\documentclass[twocolumn,times]{aastex62}

\usepackage{mathtools}
\usepackage{xspace}

\usepackage{txfonts} 
\usepackage{microtype} 

\usepackage{savesym} 
\savesymbol{tablenum} 
\usepackage{siunitx}
\restoresymbol{SIX}{tablenum} 
\usepackage{physics}
\DeclareMathAlphabet{\mathpzc}{OT1}{pzc}{m}{it}

\def\pqset{\lbrace p,q \rbrace\xspace}

\gdef\triadcond{\VK+\VP+\VQ=0}
\gdef\triadcondmag{\VKLEN\leq\VPLEN\leq\VQLEN}

\gdef\geomterm{-1/4\,\vect{h}^*_{\Sp}(\VP)\cross\vect{h}^*_{\Sq}(\VQ) \dotproduct \vect{h}^*_{\Sk}(\VK)}

\renewcommand{\Re}{\operatorname{Re}}

\newcommand{\eqnref}[1]{(\ref{#1})} 
\gdef\vect#1{\mathbf{#1}} 

\gdef\cmplxconj{\mathrm{c.c.}}

\gdef\Hmag{H_{\mathrm{m}}}
\gdef\Hcro{H_{\mathrm{c}}}

\gdef\meanhelicityu{\delta^u}
\gdef\meanhelicityB{\delta^B}
\gdef\meanhelicitycro{\gamma}
\gdef\meanhelicitycrou{\gamma^u}
\gdef\meanhelicitycroB{\gamma^B}
\gdef\meanenergy{\varepsilon}
\gdef\meanenergyu{\varepsilon^u}
\gdef\meanenergyB{\varepsilon^B}

\gdef\sprime{^{\prime}\mkern-1.2mu}
\gdef\dprime{^{\prime\prime}\mkern-1.2mu}

\gdef\Sk{s}
\gdef\Sp{s\sprime}
\gdef\Sq{s\dprime}

\gdef\VK{\vect{k}}
\gdef\VP{\vect{k\sprime}}
\gdef\VQ{\vect{k\dprime}}

\gdef\VKLEN{k}
\gdef\VPLEN{k\sprime}
\gdef\VQLEN{k\dprime}

\gdef\angp{\theta\sprime}
\gdef\angq{\theta\dprime}

\gdef\triadset{\lbrace\VK,\VP,\VQ\rbrace}

\gdef\fldplaceholderA{X}
\gdef\fldplaceholderB{Y}
\gdef\fldplaceholderC{Z}
\gdef\fldplaceholderABC{\fldplaceholderA\fldplaceholderB\fldplaceholderC}
\gdef\fldplaceholderVALID{\lbrace uuu, uBB, BuB, BBu\rbrace}

\gdef\HDSMint#1#2{I^{#1}_{n;i}\left(#2\right)}
\gdef\HDSMintshort#1{I^{#1}_{n;i}}
\gdef\MHDSMint#1#2{\HDSMint{#1}{#2}}
\gdef\Wgfunc#1#2{g_{i}}

\gdef\Wepsfunc#1#2{\epsilon_{#1}^{#2}}
\gdef\Wepsfunc#1#2{\epsilon_{i}}
\gdef\Weps{\Wepsfunc{p,q}{\Spi,\Sqi}}

\gdef\Wepsdeffuncflat#1#2#3#4{ (1 - #4\lambda^{#2})/(\lambda^{#1}-#3#4\lambda^{#2})}
\gdef\Wepsdef{\Wepsdeffuncflat{p}{q}{\Spi}{\Sqi}}

\gdef\Wxifunc#1#2{\xi_{i}}
\gdef\Wxi{\Wxifunc{p,q}{\Spi,\Sqi}}

\gdef\Wxideffunc#1#2#3{-#3(1- #2\Weps)}
\gdef\Wxidef{\Wxideffunc{p,q}{\Spi}{\Sqi}}

\gdef\Wkapfunc#1#2{\tau_{i}}
\gdef\Wkap{\Wkapfunc{p,q}{\Spi,\Sqi}}
\gdef\WkapsigB{\Wkapfunc{p,q}{\Sigp,\Sigq}}

\gdef\Wkapsig{\WkapsigB}
\gdef\Wkapdeffunc#1#2#3{1/(#2\lambda^p - #3\lambda^q)}
\gdef\Wkapdef{\Wkapdeffunc{p,q}{\Spi}{\Sqi}}

\gdef\SMtermA#1#2#3#4#5{#1_{n+p}^{#3\cdot#4,*}#2_{n+q}^{#3\cdot#5}}
\gdef\SMtermB#1#2#3#4#5{#1_{n-p}^{#3\cdot#4,*}#2_{n-p+q}^{#3\cdot#4\cdot#5}}
\gdef\SMtermC#1#2#3#4#5{#1_{n-q}^{#3\cdot#5}  #2_{n-q+p}^{#3\cdot#4\cdot#5}}

\gdef\SMfrckin#1{f_{n}^{u^{#1}}}
\gdef\SMfrcmag#1{f_{n}^{B^{#1}}}

\gdef\Guuusymb{\mathrm{G}^{uuu}}
\gdef\GuBBsymb{\mathrm{G}^{uBB}}
\gdef\GBuBsymb{\mathrm{G}^{BuB}}
\gdef\GBBusymb{\mathrm{G}^{BBu}}
\gdef\Guuu{\Guuusymb_{i}}
\gdef\GuBB{\GuBBsymb_{i}}
\gdef\GBuB{\GBuBsymb_{i}}
\gdef\GBBu{\GBBusymb_{i}}
\gdef\Gxyzi#1{\mathrm{G}^{XYZ}_{#1}}
\gdef\Guuui#1{\Guuusymb_{#1}}
\gdef\GuBBj#1{\GuBBsymb_{#1}}
\gdef\GBuBl#1{\GBuBsymb_{#1}}
\gdef\GBBum#1{\GBBusymb_{#1}}

\gdef\Pigeneral#1#2{\Pi^{#2}_{#1}(k_n)}
\gdef\Piuuui#1{\Pi^{uuu}_{#1}(k_n)}
\gdef\PiuBBj#1{\Pi^{uBB}_{#1}(k_n)}
\gdef\PiBuBl#1{\Pi^{BuB}_{#1}(k_n)}
\gdef\PiBBum#1{\Pi^{BBu}_{#1}(k_n)}
\gdef\Piuuu{\Piuuui{i}}
\gdef\PiuBB{\PiuBBj{i}}
\gdef\PiBuB{\PiBuBl{i}}
\gdef\PiBBu{\PiBBum{i}}

\gdef\Delgen#1#2#3{{\Delta_{#1#2}^{#3}}}

\gdef\Gsigns{\lbrace\Sk,\Sp,\Sq\rbrace}

\gdef\GsignsA{\pm\lbrace +,-,+\rbrace}
\gdef\GsignsB{\pm\lbrace +,-,-\rbrace}
\gdef\GsignsC{\pm\lbrace +,+,-\rbrace}
\gdef\GsignsD{\pm\lbrace +,+,+\rbrace}

\gdef\Gsignsii#1{\lbrace\Spii{#1},\Sqii{#1}\rbrace}
\gdef\GsignsAi{\lbrace -,+\rbrace}
\gdef\GsignsBi{\lbrace -,-\rbrace}
\gdef\GsignsCi{\lbrace +,-\rbrace}
\gdef\GsignsDi{\lbrace +,+\rbrace}

\gdef\alphasymb{\alpha}

\gdef\alphaBuB{\alphasymb^{BuB}}
\gdef\alphaBBu{\alphasymb^{BBu}}

\gdef\ssym{m}
\gdef\fluxsumA{\sum^{n+q}_{\ssym=n+1}}
\gdef\fluxsumB{\sum^{n+q-p}_{\ssym=n+1}}

\gdef\Sigk{\sigma}
\gdef\Sigp{\sigma\sprime}
\gdef\Sigq{\sigma\dprime}

\definecolor{uuu}{HTML}{000000}
\definecolor{uBB}{HTML}{377eb8}
\definecolor{BuB}{HTML}{ff7f00}
\definecolor{BBu}{HTML}{e41a1c}

\gdef\CA#1{\textcolor{uuu}{#1}}
\gdef\CB#1{\textcolor{uBB}{#1}}
\gdef\CC#1{\textcolor{BuB}{#1}}
\gdef\CD#1{\textcolor{BBu}{#1}}

\gdef\CA#1{#1}
\gdef\CB#1{#1}
\gdef\CC#1{#1}
\gdef\CD#1{#1}

\gdef\VKr{}
\gdef\VPr{\frac{k'}{k}} 
\gdef\VQr{\frac{k''}{k}}
\gdef\VPril{\VPLEN/\VKLEN} 
\gdef\VQril{\VQLEN/\VKLEN}

\gdef\Sk{s}
\gdef\Sp{s'}
\gdef\Sq{s''}
\gdef\Spi{s'_{i}}
\gdef\Sqi{s''_{i}}
\gdef\Spii#1{s'_{#1}}
\gdef\Sqii#1{s''_{#1}}
\gdef\Rk{\Sigk}
\gdef\Rp{\Sigp}
\gdef\Rq{\Sigq}

\gdef\dt{d_t}

\gdef\Pra{\left(\VPr\right)^\alpha}
\gdef\Qra{\left(\VQr\right)^\alpha}
\gdef\Krao{}
\gdef\Prao{\left(\VPr\right)^{\alpha+1}}
\gdef\Qrao{\left(\VQr\right)^{\alpha+1}}

\gdef\Ea{E^{(\alpha)}}

\gdef\Eka{E^{(\alpha)}(\VK)}
\gdef\EkaVX#1{E^{(\alpha)}(#1)}

\gdef\fld#1#2#3#4{#1_{#2}^{#3}}

\gdef\uk#1{\fld{u}{\Sk}{#1}{\VK}}
\gdef\up#1{\fld{u}{\Sp}{#1}{\VP}}
\gdef\uq#1{\fld{u}{\Sq}{#1}{\VQ}}

\gdef\Bk#1{\fld{B}{\Rk}{#1}{\VK}}
\gdef\Bp#1{\fld{B}{\Rp}{#1}{\VP}}
\gdef\Bq#1{\fld{B}{\Rq}{#1}{\VQ}}

\gdef\tripcorruuu{\up{*}\up{*}\uq{*}}
\gdef\tripcorruBB{\uk{*}\Bp{*}\Bq{*}}
\gdef\tripcorrBuB{\Bk{*}\up{*}\Bq{*}}
\gdef\tripcorrBBu{\Bk{*}\Bp{*}\uq{*}}

\gdef\Delu{  \CA{\g{\Sk\Sp\Sq}\up{*}\up{*}\uq{*}} }
\gdef\DelBa{ \CB{\g{\Sk\Rp\Rq}\uk{*}\Bp{*}\Bq{*}} }
\gdef\DelBb{ \CC{\g{\Rk\Sp\Rq}\Bk{*}\up{*}\Bq{*}} }
\gdef\DelBc{ \CD{\g{\Rk\Rp\Sq}\Bk{*}\Bp{*}\uq{*}} }

\gdef\g#1{g_{#1}}
\gdef\paruA{(\Sp\VPLEN-\Sq\VQLEN)}
\gdef\paruB{(\Sq\VQLEN-\Sk\VKLEN)}
\gdef\paruC{(\Sk\VKLEN-\Sp\VPLEN)}
\gdef\parBA{(\Rp\VPLEN-\Rq\VQLEN)}
\gdef\parBB{(\Rq\VQLEN-\Rk\VKLEN)}
\gdef\parBC{(\Rk\VKLEN-\Rp\VPLEN)}
\gdef\paruAr{\left(\Sp\VPr-\Sq\VQr \right)}
\gdef\paruBr{\left(\Sq\VQr-\Sk\VKr \right)}
\gdef\paruCr{\left(\Sk\VKr-\Sp\VPr \right)}
\gdef\parBAr{\left(\Rp\VPr-\Rq\VQr \right)}
\gdef\parBBr{\left(\Rq\VQr-\Rk\VKr \right)}

\gdef\parBCrrev{\left(\Rp\VPr-\Rk\VKr \right)}

\gdef\ub{\mathbf{u}}
\gdef\Bb{\mathbf{B}}
\gdef\eqnref#1{(\ref{#1})}
\gdef\defEdens{\frac{1}{2}\left(\left\lvert u_{\Sk}(\VK)\right\rvert^2 + \left\lvert B_{\Sk}(\VK)\right\rvert^2\right)}
\gdef\defHmdens{     \frac{\Sk}{\VKLEN} \left\lvert B_{\Sk}(\VK)\right\rvert^2}

\gdef\defHcdens{\Re\left[u_{\Sk}(\VK)B^*_{\Sk}(\VK)\right]}
\gdef\sumk{\sum_{\VK}}
\gdef\sums{\sum_{\Sk}}
\gdef\VKdep{(\VK)}
\gdef\VPdep{(\VP)}
\gdef\VQdep{(\VQ)}
\gdef\VKdep{}
\gdef\VPdep{}
\gdef\VQdep{}
\gdef\Fsymb{L}
\gdef\GBuBconserves{$\GBuBl{2}$ and $\GBuBl{3}$\xspace}
\gdef\GBBuconserves{$\GBBum{3}$ and $\GBBum{4}$\xspace}
\gdef\GBuBconservescompl{$\GBuBl{1}$ and $\GBuBl{4}$\xspace}
\gdef\GBBuconservescompl{$\GBBum{1}$ and $\GBBum{2}$\xspace}
\gdef\fA{\fldplaceholderA}
\gdef\fB{\fldplaceholderB}
\gdef\fC{\fldplaceholderC}
\gdef\IfB{V}
\gdef\IfC{W}
\gdef\triplekaparg{\Wkapsig,\Wkapsig,\Wkapsig}
\gdef\fABC{\fA\fB\fC}
\gdef\dtnl{d_t\vert_{\mathrm{n.l.}}}

\received{---}
\revised{---}
\accepted{---}
\submitjournal{ApJ}

\shorttitle{Partial invariants, large-scale dynamo action, and the inverse transfer of magnetic helicity}
\shortauthors{Rathmann and Ditlevsen}

\begin{document}

\title{\replaced{}{Partial invariants, large-scale dynamo action, and the inverse transfer of magnetic helicity}}

\correspondingauthor{Nicholas M. Rathmann}
\email{rathmann@nbi.ku.dk}
\author[0000-0001-7140-0931]{Nicholas M. Rathmann}
\affiliation{Niels Bohr Institute, University of Copenhagen, Denmark}
\author[0000-0003-2120-7732]{Peter D. Ditlevsen}
\affiliation{Niels Bohr Institute, University of Copenhagen, Denmark}
	
\begin{abstract}
\replaced{}{
The existence of partially conserved enstrophy-like quantities is conjectured to cause inverse energy transfers to develop embedded in magnetohydrodynamical (MHD) turbulence, in analogy to the influence of enstrophy in two-dimensional nonconducting turbulence.
By decomposing the velocity and magnetic fields in spectral space onto helical modes, we identify subsets of three-wave (triad) interactions conserving two new enstrophy-like quantities which can be mapped to triad interactions recently identified with facilitating large-scale $\alpha$-type dynamo action and the inverse transfer of magnetic helicity. Due to their dependence on interaction scale locality, the invariants suggest that the inverse transfer of magnetic helicity might be facilitated by both local- and nonlocal-scale interactions, and is a process more local than the $\alpha$-dynamo.
We test the predicted embedded (partial) energy fluxes by constructing a shell model (reduced wave-space model) of the minimal set of triad interactions (MTI) required to conserve the ideal MHD invariants.
Numerically simulated MTIs demonstrate that, for a range of forcing configurations, the partial invariants are, with some exceptions, indeed useful for understanding the embedded contributions to the total spectral energy flux. 
Furthermore, we demonstrate that strictly inverse energy transfers may develop if enstrophy-like conserving interactions are favoured, a mechanism recently attributed to the energy cascade reversals found in nonconducting three-dimensional turbulence subject to strong rotation or confinement. 
The presented results have implications for the understanding of the physical mechanisms behind large-scale dynamo action and the inverse transfer of magnetic helicity, processes thought to be central to large-scale magnetic structure formation.
}
\end{abstract}

\keywords{
magnetohydrodynamics (MHD) --- magnetic fields --- dynamo --- turbulence 
}

\section{Introduction}
\label{sec:intro}
\replaced{}{
A central problem in astrophysics is understanding how large-scale magnetic fields are generated by astrophysical bodies such as planets, stars, and galaxies \citep{parker1979cosmical}.
A popular explanation is that large-scale dynamo action occurs, which allows a small magnetic seed field to grow by stretching, twisting, and folding through interactions with the velocity field inside the electroconducting fluid of the body, whereby kinetic energy is converted to magnetic energy \citep{bib:moffatt1978field,krause1980mean,bib:brandenburg2005astrophysical,tobias2013ten}. 
A celebrated example of a large-scale dynamo is the $\alpha$-effect of mean-field electrodynamics \citep{steenbeck1966berechnung,bib:moffatt1978field,krause1980mean}, where the mean electromotive force caused by small-scale field fluctuations is related to the large-scale magnetic field by a coefficient, $\alpha$.
In the presence of small-scale kinetic helicity (net imbalance between left- and right-handed helical motion), the $\alpha$-effect leads to the development of large- and small-scale magnetic helicity of opposite signs, where the small-scale magnetic and kinetic helicity share signs \citep{bib:brandenburg2005astrophysical}. 
This process can conceptually be related to a stretch-twist-fold dynamo for closed magnetic flux tubes which generates opposite signs of magnetic helicity at large and small scales \citep{vainshtein1972reviews,childress1995stretch}.
Combining the $\alpha$-effect with the influence of differential rotation ($\omega$-effect), the $\alpha$--$\omega$ dynamo \citep{bib:moffatt1978field,parker1979cosmical,krause1980mean} has widely been invoked to explain the amplification and maintenance of large-scale magnetic fields.
In spiral galaxies, for example, this mechanism leads to predicted magnetic (spiral) pitch angles in the middle of observed ranges \citep{widrow2002origin,bib:brandenburg2005astrophysical}, among other observations in support of dynamo action \citep{shukurov2004introduction}.
Likewise, large-scale solar magnetic phenomena, such as solar flares and spots, are generally attributed dynamo action in combination with differential rotation within the solar convective zone \citep{hood2011solar,brun2017magnetism}.

Magnetic helicity is an inviscid integral of motion in magnetohydrodynamical (MHD) turbulence, and is observed in, e.g. the solar photosphere \citep{demoulin2007recent,blackman2016magnetic} and the solar wind \citep{howes2010interpretation}, and plays a role in coronal mass ejections by effecting magnetic flux tubes topologies \citep{malapaka2013modeling}.
The inverse (upscale) transfer of magnetic helicity \citep{bib:frisch1975possibility} is another transfer process suggested to contribute to large-scale magnetic structure formation by virtue of the spectral bounds between magnetic energy and magnetic helicity \citep{bib:pouquet1976strong,bib:moffatt1978field,bib:biskamp1993nonlinear}.
In spite of receiving a lot of attention, less is known about the nonlinear dynamics enabling an inverse transfer of magnetic helicity.

Simulations of homogeneous MHD turbulence in a box with triple periodic boundaries have been the subject of many studies attempting to better understand the conditions under which large-scale magnetic structure formation takes place due to the $\alpha$-effect \citep{brandenburg2001inverse,bib:linkmann2017effects} and the inverse transfer of magnetic helicity \citep{alexakis2006inverse,muller2012inverse,malapaka2013large,linkmann2016large,bib:linkmann2017triad, bib:linkmann2017effects}.
As an outcome, different degrees of scale locality among interactions between fields have been reported, and it is currently thought that long-range interactions might be more important in MHD turbulence than in nonconducting fluids \citep{mininni2011scale}.
On this note, we shall refrain ourselves from referring to the inverse transfer of magnetic helicity as a cascade process, since the latter is generally associated with a constant flux through wavenumber space due to scale-local interactions, which the transfer of magnetic helicity might not be \citep{alexakis2006inverse,aluie2010scale,muller2012inverse}.

In an attempt to better understand the mechanisms which facilitate large-scale magnetic structure formation of astrophysical interest, such as the inverse transfer of magnetic helicity and large-scale dynamo action, it is therefore important to study the nonlinear dynamics by which inviscid invariants are transferred across spatial scales. This is the focus of our work. 
}

\subsection*{Role of inviscid invariants}
In three-dimensional (3D), isotropic, hydrodynamical (HD) turbulence, kinetic energy is, on average, transferred from the large integral (pumping) scale of motion to the small, dissipative Kolmogorov scale (where dissipated as heat) by scale-local interactions, called a forward or direct energy cascade. 
In certain cases of HD turbulence, however, such as two-dimensional (2D) flows \citep{bib:boffetta2010evidence,bib:mininni2013inverse} and strongly rotating 3D flows with a broken mirror symmetry \citep{bib:sulem1989generation,bib:mininni2009scale}, an inverse (or reverse) energy cascade has been observed.

\replaced{}{
In 3D HD turbulence, the dissipation of energy at the Kolmogorov scale is proportional to the enstrophy (vorticity squared) at this scale. The energy cascade in high Reynolds number turbulence must therefore be facilitated by a production of enstrophy, which is possible by means of the stretching and bending term in the vorticity equation. 
In 2D HD turbulence, however, the stretching and bending term is absent, and enstrophy, too, is an inviscid invariant along with energy and can only grow by increased pumping.
Because the energy spectrum, $E(k)$, and the enstrophy spectrum, $Z(k)$, are related by $Z(k) = k^2 E(k)$, the cascades of the two quantities cannot be treated independently, which leads to dual and counter-directional cascades whereby enstrophy cascades forwardly and energy cascades inversely \citep{bib:kraichnan1967inertial,alexakis2018cascades}.
}

In 3D HD flows, a second inviscid invariant also exists: kinetic helicity, defined as the integral of the inner product between velocity and vorticity \citep{bib:moffatt1969degree,bib:kraichnan1973helical,bib:brissaud1973helicity}.
In contrast to enstrophy in 2D, the effect of helicity on the directionality of the energy cascade in 3D is less understood. 
Although the helicity spectrum can also dominate over the energy spectrum at small scales (large wavenumber, $k$) as enstrophy, helicity is not sign definite as opposed to enstrophy. As a consequence, helicity does not place  similar restrictions on the direction of the energy cascade \citep{alexakis2018cascades}.

By decomposing the velocity field in spectral space onto helical modes, each velocity component evolves according to the Navier--Stokes equation by helical three-wave (triad) interactions which separately conserve kinetic energy and kinetic helicity \citep{bib:constantin1988beltrami,bib:waleffe1992nature}.
Recently, new additional quantities were identified that are partially conserved among helical triad interactions \citep{bib:DePietro2015,bib:rathmannditlevsen2016,bib:rathmann2017pseudo}, henceforth referred to as \textit{pseudo-invariants} or \textit{partial invariants}; that is, quantities which are conserved only by a subset of all possible helical triad interactions.
\replaced{}{
In a further subset of helical triad interactions, the associated pseudo-invariants become enstrophy-like and have been suggested to induce embedded (partial) inverse energy cascades in 3D HD turbulence \citep{bib:rathmann2017pseudo} [relevant to the interpretation of other numerical studies such as \citet{bib:biferale2012inverse,bib:DePietro2015,bib:alexakis2016helically,bib:sahoo2017discontinuous}]. }
Since these triad interactions are predominantly responsible for channelling energy upscale within rotating flows \citep{buzzicotti2018energy}, and with possible relevance for thin-layered turbulence \citep{bib:benavides2017critical}, there are reasons to believe that inverse energy transfers might generally exist embedded in 3D HD turbulence due to partial invariants with implications for the net transfer of energy. 

\added{
In ideal MHD turbulence, the existence of three inviscid invariants involving the velocity and magnetic fields, and the existence of separate dissipation scales associated with kinematic viscosity and magnetic resistivity, make understanding the factors controlling transfer processes particularly challenging. 
For example, in the nonlinear regime where the back-reaction on the velocity field from the Lorentz force is nonnegligible, the inverse transfer of magnetic helicity may be more or less local in scale depending on the relative signs of the small-scale kinetic helicity and magnetic helicity content \citep{bib:linkmann2017effects}. 
Inverse spectral transfers can, however, occur even for vanishing kinetic and magnetic helicity \citep{brandenburg2015nonhelical}, and \citet{aluie2017coarse} showed that the transfer directionality of magnetic helicity depends on the magnetic Reynolds number (high values excluding a forward transfer).
Meanwhile, \citet{bian2019decoupled} recently demonstrated the existence of a subrange within the inertial--inductive range in which the total energy cascade decouples into separate, conservative cascades of kinetic and magnetic energy.
}

Because of their clear physical interpretation, quadratic invariants, such as enstrophy, play a central role in the study of turbulent cascade dynamics.
In this work, we conjecture that the spectral--helically decomposed energy fluxes in ideal MHD turbulence might also be understood in terms of pseudo-invariants, which has potential implications for large-scale magnetic structure formation insofar as the aggregate of triads interactions conserving them are relevant for the velocity and magnetic field evolutions.
We show that two new enstrophy-like quantities are partially conserved by the ideal MHD equations, and argue that embedded, inverse energy transfers might develop as a result.
Intriguingly, the new quantities are conserved by triad interactions recently argued to facilitate large-scale $\alpha$-type dynamo action and the inverse transfer of magnetic helicity \citep{bib:linkmann2016helical}.
By constructing a shell model (reduced wave-space model), we show the new pseudo-invariants are indeed useful for understanding the simulated partial (forward and inverse) spectral energy fluxes and, moreover, demonstrate that strictly inverse energy transfers might develop if enstrophy-like conserving interactions are favoured, such as results for nonconducting turbulence subject to strong rotation \citep{buzzicotti2018energy} or confinement \citep{bib:benavides2017critical} suggest.

\section{The spectral--helical decomposition}
 
\begin{figure*}[!t]
\centering
\includegraphics[width=0.85\textwidth]{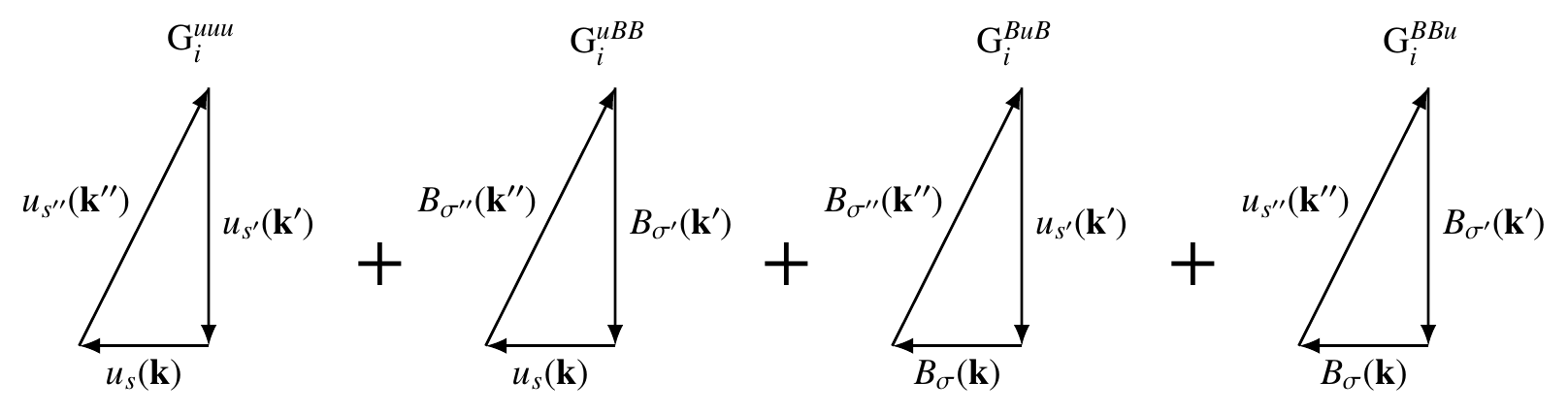} 
\caption{
Minimal set of helical triad interactions (MTI) required to conserve all three magnetohydrodynamical invariants. 
For a given triad of waves, $\triadset$, the MTI consists of four components per possible combination of helical signs of the interacting modes $\lbrace \Sk,\Sp,\Sq,\Sigk,\Sigp,\Sigq \rbrace$ (or sign group numbers $i$; see the text): one velocity triad interaction ($\Guuu$) and three velocity--magnetic triad interactions ($\GuBB,\GBuB,\GBBu$). 
The behaviour of the MTI components is proposed to be explained by the conservation (or lack thereof) of a pseudo-invariant quantity, $\Ea$.
\label{fig:MTI}}
\end{figure*}

In real space, the incompressible, ideal MHD equations are 
\begin{align}
(\partial_t-  \nu \laplacian)\ub &= -(\ub\dotproduct \grad)\ub +  (\curl \Bb )\cross\Bb  - \grad p \nonumber\\
(\partial_t- \eta \laplacian)\Bb &=  (\Bb\dotproduct \grad)\ub -  (\ub\dotproduct \grad)\Bb  \label{eqn:MHDrealspace}\\
\div\ub = 0 &\quad \text{and}\quad \div\Bb = 0 \nonumber,
\end{align}
where $\ub$ is the velocity field, $\Bb$ the magnetic field in Alfv\'{e}n units, $\nu$ is the kinematic viscosity, $\eta$ the magnetic resistivity, $p$ the pressure, and the density set to $\rho=1$ for convenience.
In spectral space, the divergence-free constraints on $\vect{u}$ and $\vect{B}$ translate into $\VK\dotproduct\vect{u}(\VK)=0$ and $\VK\dotproduct\vect{B}(\VK)=0$. The helical basis \citep{bib:constantin1988beltrami,bib:waleffe1992nature} exploits this property by decomposing each complex spectral component into two helical modes, $\vect{h}_\pm(\VK)$, which are mutually perpendicular to $\VK$ and are eigenmodes of the curl operator, i.e. $i \VK \cross \vect{h}_\pm = \pm k \vect{h}_\pm$ where $k=\vert\VK\vert$.  
In this basis, the velocity and magnetic components are given by 
\begin{align*}
\vect{u}(\VK) &= u_+(\VK)\vect{h}_+ + u_-(\VK)\vect{h}_- \\
\vect{B}(\VK) &= B_+(\VK)\vect{h}_+ + B_-(\VK)\vect{h}_-,
\end{align*}
and the ideal MHD invariants energy ($E$), magnetic helicity ($\Hmag$), and cross helicity ($\Hcro$), are given by
\begin{align}
E 		&= \sumk E(\VK)		= 	\sumk\sums \defEdens \label{eqn:defEmhd}\\
\Hmag 	&= \sumk \Hmag(\VK) =	\sumk\sums \defHmdens \label{eqn:defHm} \\
\Hcro 	&= \sumk \Hcro(\VK) =	\sumk\sums \defHcdens \label{eqn:defHc}
,
\end{align}
where $\Sk=\pm1$ is a helical sign coefficient, and $E(\VK)$, $\Hmag(\VK)$, and $\Hcro(\VK)$, are the respective spectra.

\citet{bib:lessinnes2009helical} first proposed applying the decomposition also to the MHD equations \eqnref{eqn:MHDrealspace}, giving
\begin{align}
&\left(\partial_t +\nu k^2\right) u_{\Sk}(\VK) =  \label{eqn:MHDa}\\  &\quad \sum_{\triadcond} 
\Bigg(
\sum_{\Sp,\Sq} (\Sp\VPLEN - \Sq\VQLEN) \,g_{\Sk\Sp\Sq}\,u^*_{\Sp}(\VP) u^*_{\Sq}(\VQ)
\nonumber\\
&\qquad\qquad-\sum_{\Sigp,\Sigq} (\Sigp\VPLEN - \Sigq\VQLEN) \,g_{\Sk\Sigp\Sigq}\, B^*_{\Sigp}(\VP) B^*_{\Sigq}(\VQ)
\Bigg)
\nonumber
\\[0.0em]
&\left(\partial_t+\eta k^2\right) B_{\Sigk}(\VK) = \Sigk\VKLEN \sum_{\triadcond} \Bigg(
\label{eqn:MHDb}\\ 
&\quad
\sum_{\Sigp,\Sq} g_{\Sigk\Sigp\Sq}\, B^*_{\Sigp}(\VP) u^*_{\Sq}(\VQ)
 -   
\sum_{\Sp,\Sigq} g_{\Sigk\Sp\Sigq}\, u^*_{\Sp}(\VP) B^*_{\Sigq}(\VQ)
\Bigg)
\nonumber
,
\end{align}
where $\lbrace \Sk,\Sp,\Sq,\Sigk,\Sigp,\Sigq \rbrace = \pm1$ are helical sign coefficients of the interacting helical modes, and 
$$g_{\Sk\Sp\Sq}=\geomterm$$
is an interaction coefficient.
Velocity modes $u_{\Sk}(\VK)$ thus evolve by helical triad (three-wave) interactions involving 
$\lbrace u_{\Sp}(\VP),u_{\Sq}(\VQ)\rbrace$ and 
$\lbrace B_{\Sigp}(\VP),B_{\Sigq}(\VQ)\rbrace$, while magnetic modes $B_{\Sigk}(\VK)$ evolve by helical triad interactions involving 
$\lbrace u_{\Sp}(\VP),B_{\Sigq}(\VQ)\rbrace$ and 
$\lbrace B_{\Sigp}(\VP),u_{\Sq}(\VQ)\rbrace$, provided that triads close ($\VK+\VP+\VQ=0$).

Note that the splitting of the curl of the electromotive force $\curl(\ub\cross\Bb)$ in the induction equation into an advective term, $(\ub\dotproduct \grad)\Bb$, and dynamo (stretching) term, $(\Bb\dotproduct \grad)\ub$, is obfuscated by the spectral--helical decomposition and these terms are therefore not directly associated with the two sums in \eqnref{eqn:MHDb}.

For each of the four types of helical triad interactions, $2^3=8$ distinct combinations of helical signs are possible as indicated by the sums over helical signs in \eqnref{eqn:MHDa}--\eqnref{eqn:MHDb}. 
If sorted against shared interaction coefficients, however, only four unique sign combinations remain per triad type:
for each of the triad types 
$\lbrace u_{\Sk}(\VK),u_{\Sp}(\VP),u_{\Sq}(\VQ) \rbrace$, 
$\lbrace u_{\Sk}(\VK),B_{\Sigp}(\VP),B_{\Sigq}(\VQ)\rbrace$,
$\lbrace B_{\Sigk}(\VK),u_{\Sp}(\VP),B_{\Sigq}(\VQ)\rbrace$, and \linebreak
$\lbrace B_{\Sigk}(\VK),B_{\Sigp}(\VP),u_{\Sq}(\VQ)\rbrace$, the associated helical signs 
$\lbrace \Sk,\Sp,\Sq\rbrace$, 
$\lbrace \Sk,\Sigp,\Sigq\rbrace$, 
$\lbrace \Sigk,\Sp,\Sigq\rbrace$, and 
$\lbrace \Sigk,\Sigp,\Sq\rbrace$, respectively,
may be any one of the combinations $\GsignsA,\GsignsB,\GsignsC$, and $\GsignsD$. 
From here on, these four possible combinations of helical signs shall be referred to as sign groups $i=1,2,3,4$, respectively. 
Any particular helical triad interaction may therefore be referred to by a combination of the triad type and its sign group number, henceforth denoted compactly by $\Gxyzi{i}$. 
In this notation, $i$ refers to the sign group number, and $\fldplaceholderA,\fldplaceholderB$ and $\fldplaceholderC$ are placeholders for the fields of the interacting modes. Note that only the four types 
$\fldplaceholderABC\in\fldplaceholderVALID$ 
exist, and that $\fldplaceholderA$ is the field associated with wave-vector $\VK$, $\fldplaceholderB$ with $\VP$, and $\fldplaceholderC$ with $\VQ$.

Isolating terms in \eqnref{eqn:MHDa}--\eqnref{eqn:MHDb} involving a single triad of waves, $\triadset$, only four triad interactions remain per possible combination of helical signs, namely $\Guuu + \GuBB + \GBuB + \GBBu$, hereafter referred to as a minimal set of triad interactions (MTI) following \citet{bib:linkmann2017effects} (Figure \ref{fig:MTI}).
Note that $2^6=64$ distinct MTIs are possible, and that only in the case of homochiral MTIs ($\Sk=\Sigk$, $\Sp=\Sigp$, and $\Sq=\Sigq$) do the four MTI components share sign group numbers ($i$).

By noting the cyclic property of $\g{\Sk\Sp\Sq}$, the evolution of velocity and magnetic modes in a given MTI is governed by \citep{bib:linkmann2016helical}
\begin{alignat}{3}
&\dt\uk{}\VKdep = \paruA \CA{\g{\Sk\Sp\Sq}\up{*}\VPdep\uq{*}\VQdep} - \parBA \CB{\g{\Sk\Rp\Rq}\Bp{*}\VPdep\Bq{*}\VQdep}\nonumber\\
&\dt\up{}\VPdep = \paruB \CA{\g{\Sk\Sp\Sq}\uq{*}\VQdep\uk{*}\VKdep} - \parBB \CC{\g{\Rk\Sp\Rq}\Bq{*}\VQdep\Bk{*}\VKdep}\nonumber\\
&\dt\uq{}\VQdep = \paruC \CA{\g{\Sk\Sp\Sq}\uk{*}\VKdep\up{*}\VPdep} - \parBC \CD{\g{\Rk\Rp\Sq}\Bk{*}\VKdep\Bp{*}\VPdep}\nonumber\\
&\dt\Bk{}\VKdep = \Rk\VKLEN \left( \CD{\g{\Rk\Rp\Sq}\Bp{*}\VPdep\uq{*}\VQdep} -  \CC{\g{\Rk\Sp\Rq}\up{*}\VPdep\Bq{*}\VQdep} \right) \label{eqn:MHDtriaddyn}\\
&\dt\Bp{}\VPdep = \Rp\VPLEN \left( \CB{\g{\Sk\Rp\Rq}\Bq{*}\VQdep\uk{*}\VKdep} -  \CD{\g{\Rk\Rp\Sq}\uq{*}\VQdep\Bk{*}\VKdep} \right) \nonumber\\
&\dt\Bq{}\VQdep = \Rq\VQLEN \left( \CC{\g{\Rk\Sp\Rq}\Bk{*}\VKdep\up{*}\VPdep} -  \CB{\g{\Sk\Rp\Rq}\uk{*}\VKdep\Bp{*}\VPdep} \right) \nonumber
,
\end{alignat}
where the compact notations $\uk{}=u_{\Sk}(\VK)$, $\up{}=u_{\Sp}(\VP)$, and $\uq{}=u_{\Sq}(\VQ)$ are used (and likewise for $B$).
While the relative magnitudes of energy, magnetic helicity, and cross-helicity fluxes between the three triad legs are fixed and determined by the coefficients in \eqnref{eqn:MHDtriaddyn}, 
the magnitudes and directions of the average fluxes (to or from a leg) depend on the unknown triple-correlators $\langle\tripcorruuu\rangle+\cmplxconj, \allowbreak\langle\tripcorruBB\rangle+\cmplxconj, \allowbreak\langle\tripcorrBuB\rangle+\cmplxconj$, and $\allowbreak\langle\tripcorrBBu\rangle+\cmplxconj$.

From this simpler form of the spectral dynamics, it follows that each MTI conserves energy, magnetic helicity, and cross-helicity, by noting
\begin{align*}
d_t &[E(\VK)+E(\VP)+E(\VQ)] = 0 \\
d_t &[\Hmag(\VK)+\Hmag(\VP)+\Hmag(\VQ)] = 0 \\
d_t &[\Hcro(\VK)+\Hcro(\VP)+\Hcro(\VQ)] = 0
.
\end{align*}
The ideal MHD invariants \eqnref{eqn:defEmhd}--\eqnref{eqn:defHc} are thus conserved per triad of waves, but only collectively by the four components $\Guuu + \GuBB + \GBuB + \GBBu$ constituting a MTI (for any choice of helical signs), hence the notion of a minimal set.

\citet{bib:linkmann2016helical} recently proposed the behaviour of a given MTI might be understood from the linear stability of \eqnref{eqn:MHDtriaddyn} around trivial steady-states, inspired by a similar approach in HD turbulence \citep{bib:waleffe1992nature}. 
\citet{bib:waleffe1992nature} suggested that energy, on average, flows from the most unstable triad mode (leg) and in to the other two.
In this regard, the behaviour of a given triad may be classified as either forward (F-class) if the smallest wavenumber mode (largest scale) is linearly unstable, suggesting that energy is transferred to the two smaller scales (forward cascade), or reverse (R-class) if either the middle or largest wavenumber modes are unstable, suggesting that energy is transferred either partly or fully to larger scales (inverse cascade), respectively.
By considering the linear stability of the states \linebreak $\lbrace u^*_{\Sk},u^*_{\Sp},u^*_{\Sq}, B^*_{\Sigk},B^*_{\Sigp},B^*_{\Sigq} \rbrace = \allowbreak \lbrace u_0,0,0, B_0,0,0 \rbrace, \allowbreak\lbrace 0,u_0,0, 0,B_0,0 \rbrace, \allowbreak\lbrace 0,0,U_0, 0,0,B_0 \rbrace$, where $u_0$ and $B_0$ are constants, \citet{bib:linkmann2016helical} showed that the modes $x_i = (\uk{},\Bk{})$ evolve by $\ddot{x_i}=M_{ij}x_j$ [and similarly for $x_i=(\up{},\Bp{}),(\uq{},\Bq{})$, but with different $M_{ij}$]. 
The modal (leg) stabilities therefore depend on the existence eigenvalues for $M_{ij}$ with real, positive parts, which have complicated dependencies compared to the HD case: in addition to $M_{ij}$ depending on the helical signs of the three interacting modes, such as a stability analysis of the pure HD case also does, it moreover depends on the ratio $u_0/B_0$, and the alignment between velocity and magnetic modes (cross-helicity).

\added{
In fully developed turbulence, it is, however, not immediately clear to what extent the stability properties of isolated triads are applicable to the full network of triad interactions as represented by the Navier--Stokes equation \citep{moffatt2014note}. 
Several numerical HD studies considering both decimated (biased) and unbiased triad networks, as well as studies on thin-layer turbulence and flows subject to strong rotation, suggest meanwhile that the stability properties are indeed useful for explaining the embedded (partial) flux contributions to the total energy flux, including forward--inverse transitions of the total energy flux [see the review by \citet{alexakis2018cascades}].
Extending to the MHD case, \citet{bib:linkmann2017effects} recently considered which triad interactions might be associated with large- and small-scale dynamo action and the inverse transfer of magnetic helicity, finding numerically that the helical signature of the resulting large- and small-scale magnetic field components is consistent with the influence of dominant triad interactions according to a linear stability analysis, thus demonstrating its usefulness also in MHD turbulence.

Despite the predictive skill of triad stability analyses, attributing the predicted behaviours to physical mechanisms is still an open problem, one which this work attempts to address in terms of partially conserved quantities among triad interactions (pseudo-invariants).	
}

\section{The pseudo-invariants}
\label{sec:MHDpseudos}
\begin{figure*}[!t]
\includegraphics[width=1\textwidth]{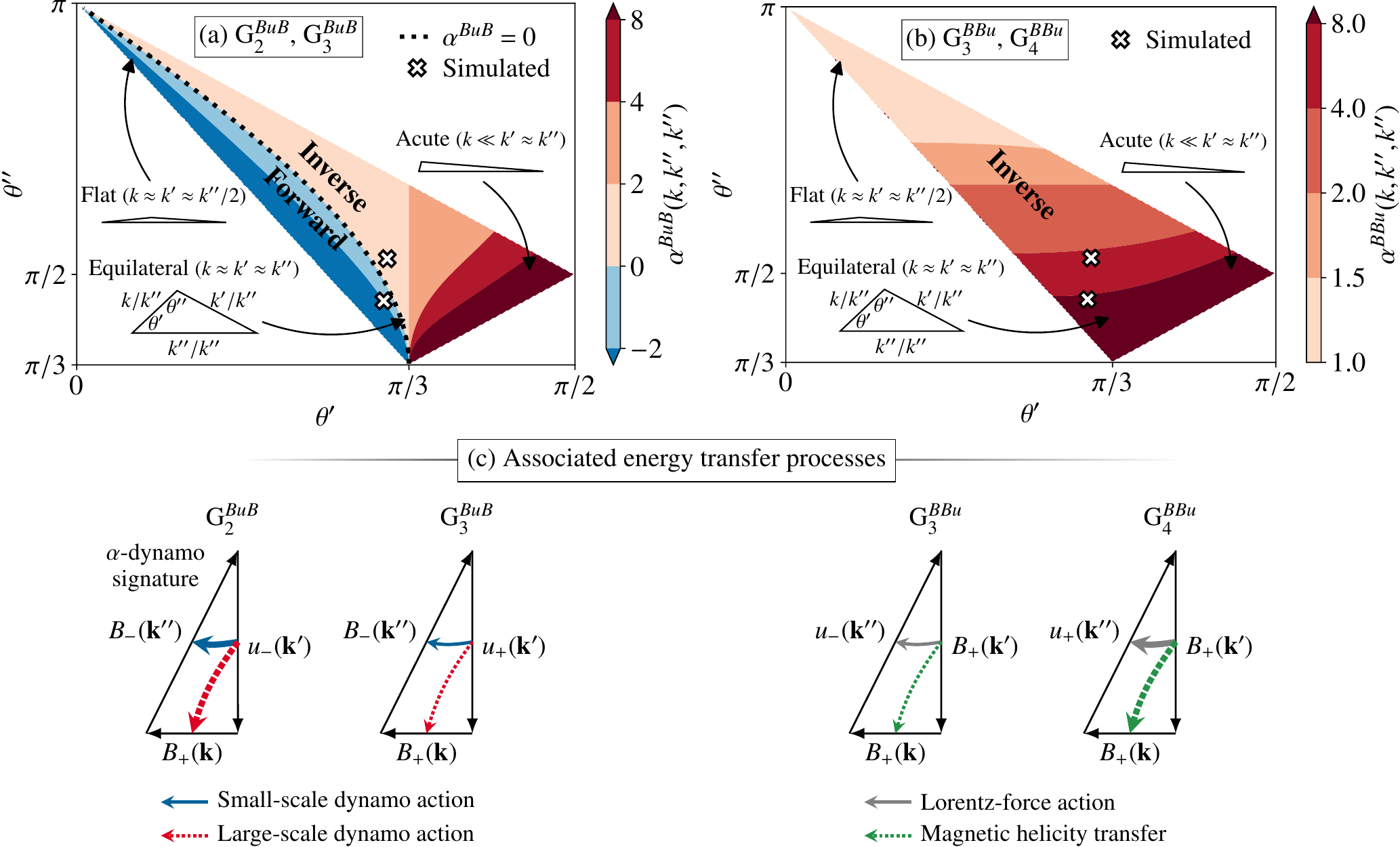} 
\caption{
Helical triad interactions conserving enstrophy-like pseudo-invariants. 
The behaviours of triad interactions $\GBuBl{2}$ and $\GBuBl{3}$ (panel \textbf{a}), and interactions $\GBBum{3}$ and $\GBBum{4}$ (panel \textbf{b}) are proposed to be explained by the conservation of the pseudo-invariant, $\Ea$, which depends on triad geometry.
Panels \textbf{a} and \textbf{b} show in red colors the subset of triad geometries conserving an enstrophy-like pseudo-invariant [exponents $\alphaBuB(\VKLEN,\VPLEN,\VQLEN)>0$ and $\alphaBBu(\VKLEN,\VPLEN,\VQLEN)>0$], which are conjectured to contribute to an inverse transfer of energy.
The black dotted line in panel \textbf{a} shows the analytical line partitioning interaction space into the triads which conserve enstrophy-like quantities and those which do not [equation \eqnref{eqn:GBuBtransition}]. Crosses in panels \textbf{a} and \textbf{b} mark the triad geometries considered in numerical simulations.
\added{
Panel \textbf{c} shows the associated energy transfer processes as identified from a linear stability analysis of \eqnref{eqn:MHDtriaddyn} \citep{bib:linkmann2016helical,bib:linkmann2017triad}: dashed arrows represent upscale transfer processes while solid arrows represent downscale transfer processes; thick and thin arrows represent the dominant and subordinate transfers, respectively.  
}
\label{fig:triadtypes_MHD}}
\end{figure*}

In HD turbulence, quadratic invariants play a fundamental role in the understanding of turbulent cascade dynamics, such the energy cascade reversal in 2D which may be understood by both energy and enstrophy being strictly positive quantities and enstrophy dominating the small scales \citep{alexakis2018cascades}. 
The fact that $\Guuui{4}$ interactions in 3D conserve both signs of kinetic helicity separately is an intriguing possible explantation for the identified R-class nature of $\Guuui{4}$ interactions \citep{bib:biferale2012inverse} in analogy to enstrophy-conserving interactions in 2D.
Recently, the mixed forward--inverse behaviour exhibited by $\Guuui{2}$ interactions depending on triad geometry \citep{bib:waleffe1992nature,bib:DePietro2015,bib:rathmann2017pseudo} was also proposed to be explained by the existence of a new enstrophy-like pseudo-invariant $\Eka = \VKLEN^{\alpha} ( \vert u_{+}(\VK)\vert^2 + \vert u_{-}(\VK)\vert^2 )$, where $\alpha \in \mathbb{R}$ depends on the specific triad shape $\lbrace\VKLEN,\VPLEN,\VQLEN\rbrace$ and hence interaction locality in wave space \citep{bib:rathmann2017pseudo}.

Following \citet{bib:rathmann2017pseudo}, we here show that a subset of MHD triad interactions exists that also conserve enstrophy-like quantities, which might help in understanding the behaviour of MTIs in terms of conserved quantities in analogy to enstrophy-conserving interactions in 2D HD turbulence.
Inspired by the HD pseudo-invariant, consider therefore the generalized spectral energy density defined as
\begin{alignat}{3}
\Eka &= \VKLEN^{\alpha} \left( \vert u_{+}(\VK)\vert^2 + \vert u_{-}(\VK)\vert^2  + \vert B_{+}(\VK)\vert^2 + \vert B_{-}(\VK)\vert^2 \right)
. 
\label{eqn:MHDpseudodef}
\end{alignat}
This quantity is thus enstrophy-like for exponents $\alpha>0$, and might induce an inverse (upscale) contribution to the total transfer of energy if conserved by triad interactions.
Note that $\alpha=0$ corresponds to energy which is conserved by any MTI.

Applying \eqnref{eqn:MHDtriaddyn} to \eqnref{eqn:MHDpseudodef}, it follows that the pseudo-invariant is conserved if $\dt [ \EkaVX{\VK}+\EkaVX{\VP}+\EkaVX{\VQ} ] = 0$, implying
\begin{align*}
 \VKLEN^{\alpha+1}\Big[ 
 &\Fsymb^{uuu}(\alpha)\,\Delu  
-\Fsymb^{uBB}(\alpha)\,\DelBa \nonumber \\
-&\Fsymb^{BuB}(\alpha)\,\DelBb 
+\Fsymb^{BBu}(\alpha)\,\DelBc
\Big] \nonumber\\ & + \cmplxconj = 0
,
\end{align*}
where [suppressing dependences on wavenumbers and helical signs in the definitions of $\Fsymb^{\fldplaceholderABC}(\alpha)$]
\begin{alignat}{2} 
&\Fsymb^{uuu}(\alpha) = \nonumber\\&\; 
\paruAr + \paruBr\Pra + \paruCr\Qra \label{eqn:Fuuu}\\
&\Fsymb^{uBB}(\alpha) = \parBAr - \Rp\Prao + \Rq\Qrao \label{eqn:FuBB}\\
&\Fsymb^{BuB}(\alpha) = \Rk\Krao + \parBBr\Pra  - \Rq\Qrao \label{eqn:FBuB}\\
&\Fsymb^{BBu}(\alpha) = \Rk\Krao - \Rp\Prao + \parBCrrev\Qra \label{eqn:FBBu}
.
\end{alignat}
Conservation by an MTI thus requires the existence of solutions $\alpha\in\mathbb{R}$ that simultaneously fulfil $\Fsymb^{\fldplaceholderABC}(\alpha) = 0$ for all $\fldplaceholderABC\in\fldplaceholderVALID$. 
Since this is not possible, we are here interested in the possibility that the four individual MTI components (Figure \ref{fig:MTI}) might separately conserve different pseudo-invariants, leading to inverse partial fluxes developing.

Equations \eqnref{eqn:Fuuu}--\eqnref{eqn:FBBu} are functions of the wavenumber ratios $\VPLEN/\VKLEN$ and $\VQLEN/\VKLEN$, and consist of a constant term and two monotonically increasing or decreasing terms. 
\added{Thus, the psuedo-invariants depend on triad shape and hence triad locality.}
For solutions $\Fsymb^{\fldplaceholderABC}(\alpha)=0$ to exist besides energy ($\alpha=0$), the signs of the three terms in each of \eqnref{eqn:Fuuu}--\eqnref{eqn:FBBu} must alternate (considering the ordering $\triadcondmag$ without loss of generality). 
On inspection, it follows that only interactions $\Gxyzi{i}\in\lbrace\Guuui{2},\GBuBl{2},\GBuBl{3},\GBBum{3},\GBBum{4}\rbrace$ solve $\Fsymb^{\fldplaceholderABC}(\alpha)=0$ for $\alpha\neq0$.

The HD pseudo-invariant associated with $\Guuui{2}$ triads has previously been investigated \citep{bib:rathmann2017pseudo}, finding a cascade reversal indeed takes place for a nonlocal subset of triad geometries due to the invariant becoming enstrophy-like ($\alpha>0$).
The pseudo-invariant associated with the velocity--magnetic triads $\GBuBl{2},\GBuBl{3},\GBBum{3}$, and $\GBBum{4}$ have, however, not previously been considered.
Unlike the HD $L$-term \eqnref{eqn:Fuuu}, the $L$-terms \eqnref{eqn:FuBB}--\eqnref{eqn:FBBu} do not depend on the helical signs of all three triad legs:  
the helical sign of the velocity mode does not enter \eqnref{eqn:FuBB}--\eqnref{eqn:FBBu}, implying \GBuBconserves share $L$-terms, and thus pseudo-invariants, as do \GBBuconserves.

Figures \ref{fig:triadtypes_MHD}a and \ref{fig:triadtypes_MHD}b show the numerical solutions w.r.t. $\alpha$ for $\Fsymb^{BuB}(\alpha)=0$ and $\Fsymb^{BBu}(\alpha)=0$, respectively, hereafter referred to as $\alphaBuB$ and $\alphaBBu$.
The solutions are shown for all possible (noncongruent) triad geometries (coloured area) by expressing each triad in terms of the two interior angles $\angp$ and $\angq$ using the Sine rule: $\VPLEN/\VKLEN = \sin(\angp)/\sin(\pi-\angp-\angq)$ and $\VQLEN/\VKLEN=\sin(\angq)/\sin(\pi-\angp-\angq)$. 
Figure \ref{fig:triadtypes_MHD}a shows that \GBuBconserves interactions may, similarly to $\Guuui{2}$, contribute either to a forward transfer of energy ($\alphaBuB<0$, blue colors) or inversely ($\alphaBuB>0$, red colors).
The exact transition line through the space of triad geometries separating the two behaviours is given by 
${d\Fsymb^{BuB}(\alpha)}/{d\alpha}\vert_{\alpha=0} = 0$, 
yielding 
\begin{align}
%
%
\log\VPril  = \frac{\log \VQril}{1-\Sigk\Sigq\VKLEN/\VQLEN}  \label{eqn:GBuBtransition}
%
,
\end{align}
which corresponds to the geometries for which the energy and pseudo-invariant solutions collapse (Figure \ref{fig:triadtypes_MHD}a, black dotted line).
Figure \ref{fig:triadtypes_MHD}b, on the other hand, shows that \GBBuconserves interactions always conserve an enstrophy-like quantity ($\alphaBBu>0$) regardless of triad geometry, suggesting they should contribute with an inverse transfer of energy.

We have thus arrived at several testable predictions for the contributions to the total transfer of energy from the four individual MTI components: 
(i) $\GuBB$ triads contribute to a forward transfer (for all sign groups $i=1,2,3,4$), 
(ii) \GBuBconservescompl contribute to a forward transfer while \GBuBconserves may contribute either to a forward or inverse transfer depending on triad geometry, and 
(iii) \GBBuconservescompl contribute to a forward transfer while \GBBuconserves contribute inversely.

\added{
\subsection{Associated energy transfer processes}

It is intriguing to note that \GBBuconserves interactions might be associated with an inverse transfer of magnetic energy since both the smallest and middle triad legs (largest and middle scale) are magnetic components. Because the large-scale magnetic components are of the same helical sign, a connection to the transfer of magnetic helicity is conceivable. Or, put differently: given a transfer of magnetic helicity (e.g. cascade), triad interactions \GBBuconserves might contribute inversely because they permit an upscale transfer of the magnetic energy associated with each helical sign due to enstrophy-like conservations (Figure \ref{fig:triadtypes_MHD}c, right-hand side).
Indeed, this behaviour has already been suggested by \citet{bib:linkmann2016helical} and \citet{bib:linkmann2017triad} who studied the stability properties of \eqnref{eqn:MHDtriaddyn} given a steady solution for the magnetic field subject to velocity and magnetic perturbations.
Note that the transfers between triad legs as indicated by the colored arrows in Figure \ref{fig:triadtypes_MHD}c follow from \citet{bib:linkmann2016helical}. 

In their work, \citet{bib:linkmann2016helical,bib:linkmann2017effects} and \citet{bib:linkmann2017triad} furthermore considered the stability properties of a steady solution for the velocity field subject to velocity and magnetic perturbations. 
In this case, contributions to the evolution of the magnetic field components comes from the velocity field, which they associated with dynamo action.
Several triad interactions were identified which may facilitate both small- and large-scale growth of magnetic field components, thus allowing the operation of small- and large-scale dynamos to be understood at the level of triad interactions.
Specifically, it was found that \GBuBconserves interactions (Figure \ref{fig:triadtypes_MHD}c, left-hand side) in the limit of nonlocal triad geometries ($\VKLEN\ll \VPLEN\approx\VQLEN$) might contribute to large-scale magnetic field growth by large-scale dynamo action, which we find is supported by the nonlocal dependence on triad geometry of the pseudo-invariant (Figure \ref{fig:triadtypes_MHD}a). (A discussion of the scale locality of interactions is deferred to section \ref{sec:scalelocality}.)

On this matter, \citet{bib:linkmann2016helical} noted that $\GBuBl{2}$ triads may produce a helical signature reminiscent of a large-scale $\alpha$-dynamo; that is, allowing for the production of large- and small-scale magnetic helicity of opposite signs from a helical velocity field such that the signs of the small-scale kinetic and magnetic helicity match \citep{bib:brandenburg2005astrophysical}.
Indeed, numerical simulations have since found support to support this identification \citet{bib:linkmann2017triad}.
If further combined with the small-scale dynamo process associated with $\GuBBj{1}$ \citep{bib:linkmann2017effects}---for which small-scale magnetic field components are amplified if their helicity matches that of the flow---the dynamo processes facilitated by $\GBuBl{2}$ triads may produce a helical signature compatible with the stretch-twist-fold dynamo \citep{vainshtein1972reviews,moffatt1989stretch,childress1995stretch,mininni2011scale} as pointed out by \citet{bib:linkmann2016helical} and \citet{bib:linkmann2017effects}. 
Note that the small-scale dynamo process associated with $\GuBBj{1}$ triads (and $\GuBBj{3}$) identified by \citet{bib:linkmann2017effects} is not at odds with the nonexistence of $\GuBBj{i}$ pseudo-invariants since a forward energy transfer is implied in that case.
}

\section{The MTI shell model} 
\label{sec:MHDshellmodel}
\begin{figure*}[!t]
\centering
\includegraphics[width=0.99\textwidth]{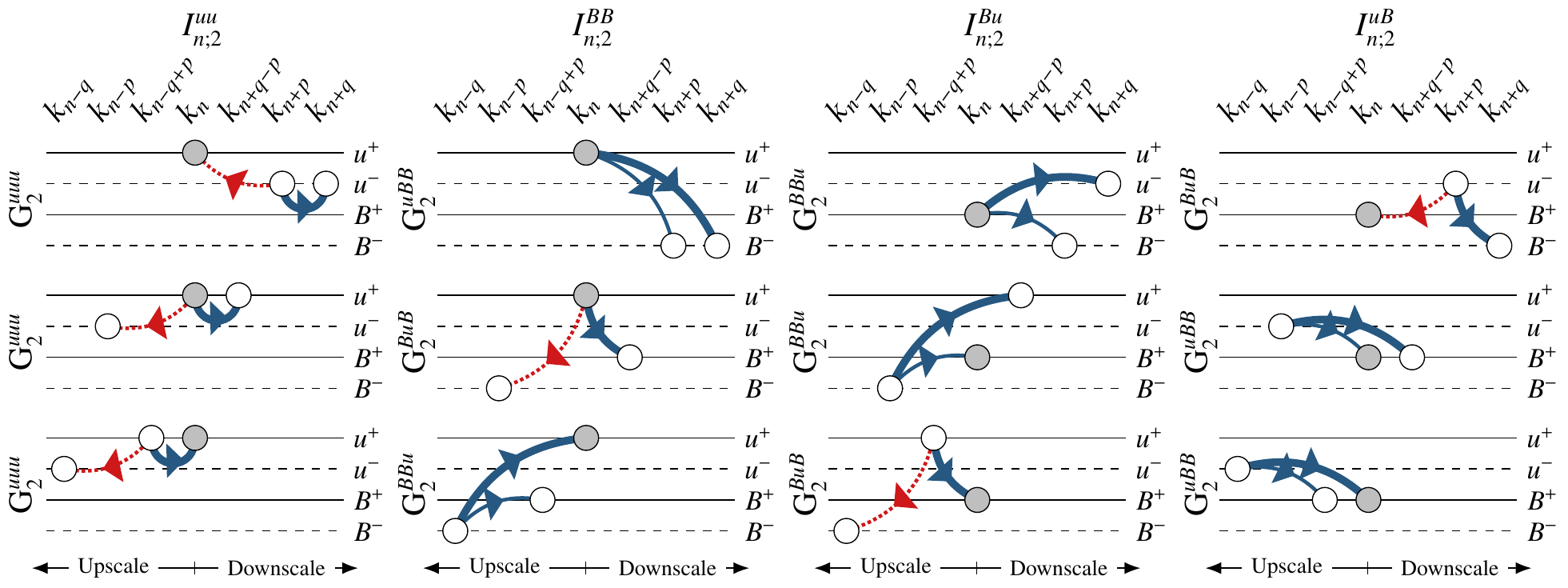} 
\caption{
The three helical triad interactions per shell model term $\HDSMintshort{uu},\HDSMintshort{BB},\HDSMintshort{Bu}$, and $\HDSMintshort{uB}$ for sign group $i=2$. 
Only interactions coupling to $u_n^+$ or $B_n^+$ (filled gray circles) are shown, complementary interactions coupling to $u_n^-$ or $B_n^-$ or are given by similar but sign-flipped interactions.
While the terms $\HDSMintshort{BB},\HDSMintshort{Bu},$ and $\HDSMintshort{uB}$ each contain a mix of the three velocity--magnetic triads ($\GuBB,\GBuB$, and $\GBBu$), the term $\HDSMintshort{uu}$ contains only velocity triads ($\Guuu$).
Arrows indicate the average energy transfer directions predicted from the pseudo-invariant conservations combined with \eqnref{eqn:MHDtriaddyn} (fixing the relative leg-to-leg transfer magnitudes): solid blue (dashed red) arrows denote forward (inverse) energy transfers, while thick (thin) arrows represent dominant (subordinate) transfers.
Note that the behaviours of $\Guuui{2}$ and $\GBuBl{2}$ interactions are shown assuming local triad geometries (dominant forward transfers).
\label{fig:sminteractions}}
\end{figure*}

To test the predictions based on the pseudo-invariants, we constructed an MTI shell model (reduced wave-space model). 
Shell models have previously provided valuable insight into the spectral dynamics of MHD turbulence \citep{bib:lessinnes2009helical,bib:plunian2013shell}, and are especially convenient in MHD turbulence due to the added number of nonlinear interactions (Figure \ref{fig:MTI}) making direct numerical simulations computationally expensive.
Only relatively recently have direct numerical simulations of \eqnref{eqn:MHDa}--\eqnref{eqn:MHDb} been conducted giving some insight into the behaviour of helical MHD triad dynamics and the role of the ideal MHD invariants in the spectral--helical basis \citep{bib:linkmann2017effects} (see section \ref{sec:discussion}).

Shell models allow simulating very long inertial--inductive ranges at the expense of severely truncating spectral space: 
only wavenumbers distributed exponentially according to $k_n=k_0\lambda^n$ are retained, where $n=0,\cdots,N$ are the shell indices, $k_0\in\mathbb{R}_+$, and $\lambda\in\;]1,(1+\sqrt{5})/2] = \;]1,\varphi]$. 
The golden ratio $\varphi$ is the upper limit such that any set of nearest neighbour waves fulfils the triangle inequality as required by \eqnref{eqn:MHDa}--\eqnref{eqn:MHDb}. 
All spectral components are generally reduced to depend only on wave magnitudes expect for a few cases \citep{gurcan2017nested,monthus2018various,gurcan2018nested}, and shell models may therefore be regarded as simple, structureless cascade models.

The pioneering work on constructing helical shell models was done by \citet{bib:benzi1996helical}, which has since inspired other helical shell models and led to important insights on helically decomposed triad dynamics of HD turbulence 
\citep{bib:ditlevsen2001dissipation,bib:ditlevsen2001cascades,bib:lessinnes2011dissipation,bib:DePietro2015,bib:rathmannditlevsen2016,bib:sahoo2017helicity,bib:DePietro2016chaotic}
and MHD turbulence  
\citep{bib:lessinnes2009helical,bib:plunian2013shell,bib:Stepanov2015}.
Following \citet{bib:rathmannditlevsen2016} for the Navier--Stokes equations, \eqnref{eqn:MHDa}--\eqnref{eqn:MHDb} may similarly be cast into a shell model by noting that the triadic interaction structure is similar but with different interaction coefficients. 
Considering only homochiral MTIs (elaborated on below) with fixed shape triads (fixed interior angles) and adopting the usual shorthand notation $u_n^{\Sk} = u_\Sk(k_n)$ and $B_n^{\Sk} = B_\Sk(k_n)$, the shell model for a single kind of MTI is given by
\begin{align}
\left(d_t+ D^u_n\right)u_n^{\Sk} &= \Sk\VKLEN_n  
\left[ \HDSMint{uu}{1,\frac{\Weps}{\lambda^p},\frac{\Wxi}{\lambda^q}} - \HDSMint{BB}{1,\frac{\Weps}{\lambda^p},\frac{\Wxi}{\lambda^q}} \right]
 + \SMfrckin{\Sk} 
\label{eqn:MHDshellmodelKin} 
\\
\left(d_t+ D^B_n\right)B_n^{\Sk} &= 
\Sk\VKLEN_n 
\left[ \MHDSMint{Bu}{\triplekaparg} - \MHDSMint{uB}{\triplekaparg} \right] + \SMfrcmag{\Sk}, 
\label{eqn:MHDshellmodelMag}
\end{align}
where the resolved triad interactions are collected in $\HDSMintshort{\IfB\IfC}$, defined as 
\begin{align*}
\HDSMint{\IfB\IfC}{a,b,c} =  
  a\SMtermA{\IfB}{\IfC}{\Sk}{\Spi}{\Sqi} 	   
- b\SMtermB{\IfB}{\IfC}{\Sk}{\Spi}{\Sqi} 		 
+ c\SMtermC{\IfB}{\IfC}{\Sk}{\Spi}{\Sqi} .
\end{align*}
Here, the helical signs of each sign group are for convenience referred to by $\Gsigns= \pm\lbrace +,\Spi,\Sqi \rbrace$, such that $\Gsignsii{i} = \GsignsAi,\GsignsBi,\GsignsCi,\GsignsDi$ for groups $i=1,2,3,4$, respectively, and the interactions coefficients are given by
\begin{align}
\Weps &= \Wepsdef \nonumber\\
\Wxi  &= \Wxidef  \nonumber\\
\Wkap &= \Wkapdef \nonumber
.
\end{align}

The integers $\pqset$ can be related to any triangular shape through the Sine rule.  
The possible resolved triad shapes depend therefore on the combination of $\lbrace \lambda, p, q\rbrace$:
For $\lambda\rightarrow 1$ any triad geometry may be constructed for large/small enough values of $\lbrace p,q\rbrace$, while for $\lbrace \lambda,p,q\rbrace = \lbrace \varphi,1,2\rbrace$ triads collapse to a line. 
Hence, for each chosen set of $\lbrace \lambda,p,q\rbrace$, the shell model consists, independently of scale $k_n$, only of fixed-shaped triad interactions.
The outer sums over all triad shapes in \eqnref{eqn:MHDa}--\eqnref{eqn:MHDb} are thus reduced to just three (fixed-shape) helical triad interactions per MTI component per resolved scale $k_n$, exemplified in Figure \ref{fig:sminteractions} for the case of $i=2$. 
Note that only $\HDSMintshort{uu}$ contains triad interactions of one MTI component exclusively, namely $\Guuu$. 
The remaining three ($\HDSMintshort{BB},\HDSMintshort{uB}$, and $\HDSMintshort{Bu}$) contain a mix of the velocity--magnetic triads $\GuBB,\GBuB$ and $\GBBu$. 

The dissipation terms are defined as $D^u_n = \nu \VKLEN_n^2+ \nu_L \VKLEN^{-4}$ and $D^B_n=\eta \VKLEN_n^2 + \eta_L \VKLEN^{-4}$ where the drag terms, $\nu_L \VKLEN^{-4}$ and $\eta_L \VKLEN^{-4}$, are added the usual way to remove energy building up at large scales.

Like the ideal MHD equations \eqnref{eqn:MHDa}--\eqnref{eqn:MHDb}, the MTI shell model also inviscidly conserves energy ($E$), magnetic helicity ($\Hmag$), and cross-helicity ($\Hcro$), which can be shown by applying \eqnref{eqn:MHDshellmodelKin}--\eqnref{eqn:MHDshellmodelMag} to \eqnref{eqn:defEmhd}--\eqnref{eqn:defHc}, telescoping sums, and inserting the boundary conditions $u_n^{\Sk},B_n^{\Sk}=0$ for $n<0$ and $n>N$.
Unlike the ideal MHD equations \eqnref{eqn:MHDa}--\eqnref{eqn:MHDb}, however, the MTI shell model \eqnref{eqn:MHDshellmodelKin}--\eqnref{eqn:MHDshellmodelMag} conserves all MHD invariants only if the resolved MTIs are homochiral; that is, each MTI component must share the same sign group, $i$. Hence, the four possible shell model MTIs are $\Guuu+\GuBB+\GBuB+\GBBu$ for $i=1,2,3,4$.

\subsection{Forcing mechanism}
The velocity ($\SMfrckin{\Sk}$) and magnetic ($\SMfrcmag{\Sigk}$) forcing terms may be constructed to allow full control over the pumping of energy ($\meanenergy=\meanenergyu+\meanenergyB$), kinetic helicity ($\meanhelicityu$), magnetic helicity ($\meanhelicityB$), and cross-helicity ($\meanhelicitycro=\meanhelicitycrou+\meanhelicitycroB$), where the superscripts $u$ and $B$ refer to the injection fields. 
For the kinetic forcing term, this amounts to solving the forcing balance equations 
$\meanenergyu      = \SMfrckin{+}u_n^{+,*} + \SMfrckin{-}u_n^{-,*} + \cmplxconj$, 
$\meanhelicityu    = k_n (\SMfrckin{+}u_n^{+,*} - \SMfrckin{-}u_n^{-,*} + \cmplxconj)$, 
$\meanhelicitycrou = \SMfrckin{+}B_n^{+,*} + \SMfrckin{-}B_n^{-,*} + \cmplxconj$, and 
$0				   = \SMfrckin{+}B_n^{+,*} - \SMfrckin{-}B_n^{-,*} + \cmplxconj$. 
For the magnetic forcing term, the balance equations are similar but with $u$ and $B$ interchanged, and with the balance of magnetic helicity being given by $\meanhelicityB = k_n^{-1}(\SMfrcmag{+}B_n^{+,*} - \SMfrcmag{-}B_n^{-,*} + \cmplxconj)$. 
Solving these closed sets of equations, the mechanical--electromagnetic helical forcing becomes
\begin{align*}
\SMfrckin{\Sk} = \frac{1}{2}\frac{(\meanenergyu+\Sk\meanhelicityu/k_n)B_n^{\Sk}-\meanhelicitycrou u_n^{\Sk}}{u_n^{\Sk,*}B_n^{\Sk} - u_n^{\Sk}B_n^{\Sk,*}}
 \\[0.6em]
\SMfrcmag{\Sk} = \frac{1}{2}\frac{(\meanenergyB+\Sk\meanhelicityB k_n)u_n^{\Sk}-\meanhelicitycroB B_n^{\Sk}}{B_n^{\Sk,*}u_n^{\Sk} - B_n^{\Sk}u_n^{\Sk,*}}
,
\end{align*}
which provides a constant pumping of energy, kinetic helicity, magnetic helicity, and cross-helicity, for constant $\meanenergy$, $\meanhelicityu$, $\meanhelicityB$, and $\meanhelicitycro$, respectively, thus allowing the simulated spectral fluxes to easily be normalized against pumping rates.

\subsection{Spectral energy flux}
The total spectral flux of energy (kinetic+magnetic), carried by a homochiral MTI of sign group $i$ through the $k_n$th scale, is given by 
$\Pigeneral{i}{} = \dtnl \sum_{m=0}^{n} \left( \vert u_m^{+}\vert^2 + \vert u_m^{-}\vert^2 + \vert B_m^{+}\vert^2 + \vert B_m^{-}\vert^2 \right)$, where $\dtnl$ includes only the nonlinear terms in \eqnref{eqn:MHDshellmodelKin}--\eqnref{eqn:MHDshellmodelMag}.
Decomposing the total flux into the four separate MTI-component contributions, or \textit{partial fluxes}, such that
\begin{align*}
\Pi_i(k_n) 
= 
\Piuuu +\PiuBB +\PiBuB+\PiBBu
,
\end{align*}
requires isolating the terms in \eqnref{eqn:MHDshellmodelKin}--\eqnref{eqn:MHDshellmodelMag} corresponding to triad interactions $\Guuu,\GuBB,\GBuB$, and $\GBBu$, which can be shown to give
\gdef\mysep{0.3em}
\begin{align*}
\Guuu &\; 
\left\{\begin{aligned}
\dtnl u_n^{\Sk} &= \Sk\VKLEN_n \HDSMint{uu}{1,\frac{\Weps}{\lambda^p},\frac{\Wxi}{\lambda^q}} \\
\dtnl B_n^{\Sk} &= 0
\end{aligned}\right.
\\[\mysep]
\GuBB &\; 
\left\{\begin{aligned}
\dtnl u_n^{\Sk} &= -\Sk\VKLEN_n \HDSMint{BB}{1,0,0} \\
\dtnl B_n^{\Sk} &= -\Sk\VKLEN_n \left[\MHDSMint{uB}{0,\Wkapsig,0} + \MHDSMint{uB}{0,0,\Wkapsig}\right]
\end{aligned}\right.
\\[\mysep]
\GBuB &\;
\left\{\begin{aligned}
\dtnl u_n^{\Sk} &= -\Sk\VKLEN_n \HDSMint{BB}{0,\frac{\Weps}{\lambda^p},0}\\
\dtnl B_n^{\Sk} &=  \Sk\VKLEN_n \left[\MHDSMint{Bu}{0,0,\Wkapsig} - \MHDSMint{uB}{\Wkapsig,0,0}\right] 
\end{aligned}\right.
\\[\mysep]
\GBBu &\;
\left\{\begin{aligned}
\dtnl u_n^{\Sk} &= -\Sk\VKLEN_n \HDSMint{BB}{0,0,\frac{\Wxi}{\lambda^q}} \\
\dtnl B_n^{\Sk} &=  \Sk\VKLEN_n \left[\MHDSMint{Bu}{\Wkapsig,0,0} + \MHDSMint{Bu}{0,\Wkapsig,0} \right]
\end{aligned}\right.
.
\end{align*}
Using these MTI-decomposed equations, the partial fluxes follow as (telescoping sums and applying the boundary conditions $u_n^{\Sk},B_n^{\Sk}=0$ for $n<0$ and $n>N$):
\gdef\mysep{0.4em}
\begin{align}
\Piuuui{i} &= \textstyle+\fluxsumA \Delgen{\ssym}{;i}{uuu} - \Sp\Weps \fluxsumB \Delgen{\ssym}{;i}{uuu} \label{eqn:SMenergyfluxA} \\[\mysep] 
\PiuBBj{i} &= \textstyle-\fluxsumA \Delgen{\ssym}{;i}{uBB} + \Sp\lambda^p\Wkapsig \fluxsumB \Delgen{\ssym}{;i}{uBB} \label{eqn:SMenergyfluxB} \\[\mysep] 
\PiBuBl{i} &= \textstyle-\Wkapsig\fluxsumA \Delgen{\ssym}{;i}{BuB} + \Sp\Weps \fluxsumB \Delgen{\ssym}{;i}{BuB} \label{eqn:SMenergyfluxC} \\[\mysep] 
\PiBBum{i} &= \textstyle+\Wkapsig\fluxsumA \Delgen{\ssym}{;i}{BBu} - \Sp\lambda^p\Wkapsig \fluxsumB \Delgen{\ssym}{;i}{BBu} \label{eqn:SMenergyfluxD}  
,
\end{align}
where the triple correlators, $\Delgen{\ssym}{;i}{\fABC}$, are defined as
$$\Delgen{\ssym}{;i}{\fABC} =\allowbreak 2k_{m-q}\Re
\left[
\fA_{m-q}^{+,*}\fB_{m-q+p}^{ \Spi,*}\fC_{m}^{ \Sqi} - \allowbreak
\fA_{m-q}^{-,*}\fB_{m-q+p}^{-\Spi,*}\fC_{m}^{-\Sqi}
\right].$$

\added{
Note that the partial fluxes \eqnref{eqn:SMenergyfluxA}--\eqnref{eqn:SMenergyfluxD} are of the total energy (kinetic+magnetic). Thus, in the same way each MTI component separately conserves total energy as follows from setting $\alpha=0$ in \eqnref{eqn:Fuuu}--\eqnref{eqn:FBBu}, the partial fluxes are based on conservative dynamics.
This, however, does not apply to the individual kinetic and magnetic energy flux contributions to the partial fluxes, and the pseudo-invariants are therefore not relevant to them.
}

\begin{figure*}[!t]
\includegraphics[width=0.99\textwidth]{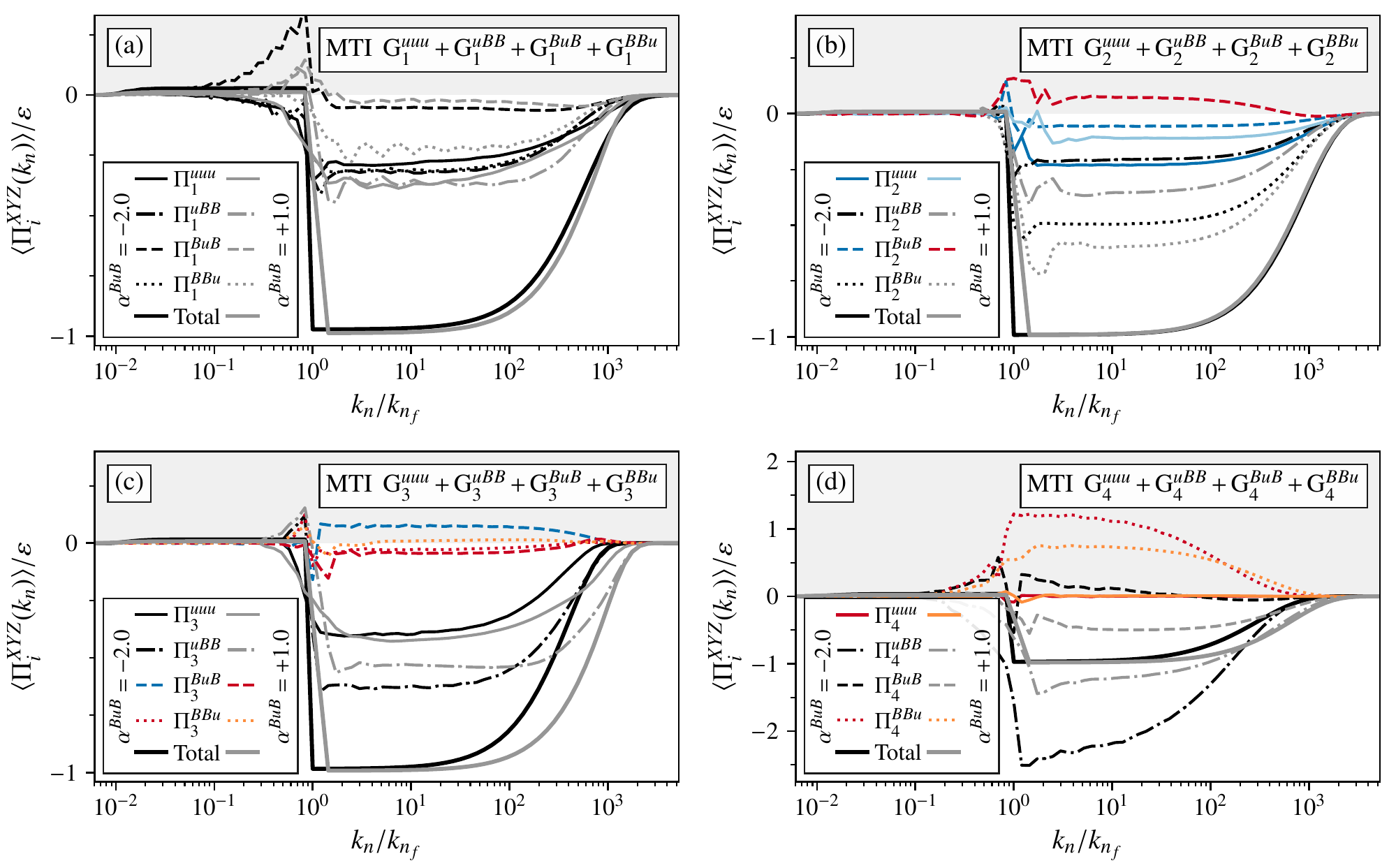} 
\caption{
Simulated partial energy fluxes associated with each MTI component for each of the four possible homochiral MTIs $\Guuu+\GuBB+\GBuB+\GBBu$ ($i=1,2,3,4$) in panels \textbf{a}--\textbf{d}, respectively.
The partial fluxes are calculated using \eqnref{eqn:SMenergyfluxA}--\eqnref{eqn:SMenergyfluxD}.
Negative (positive) values correspond to forward/downscale (inverse/upscale) fluxes. 
Blue lines denote hypothesised forward contributions according to the pseudo-invariant predictions, while red/orange lines denote hypothesised inverse contributions (interactions conserving an enstrophy-like quantity). 
Black/gray lines denote contributions by interactions not conserving any pseudo-invariants.
\label{fig:MHDcoupled}}
\end{figure*}

\section{Numerical results}
\label{sec:MHDresults}
For each of the four homochiral MTIs, two different triad shapes  
(crosses in Figure \ref{fig:triadtypes_MHD})
were considered in order to test the partial flux predictions.
The model was configured using a shell spacing of $\lambda=1.2$ with 
$\pqset=\lbrace 1,2\rbrace$ and $\pqset=\lbrace 3,4\rbrace$, corresponding to pseudo-invariant exponents of 
$\alphaBuB=\lbrace -2.0, +1.0 \rbrace$ and $\alphaBBu=\lbrace +8.8, +4.5 \rbrace$, respectively.
The chosen triad geometries thus sample contributions from both the forward and inverse parts of $\GBuB$ interaction space (Figure \ref{fig:triadtypes_MHD}a).

For simplicity, all model simulations were configured with identical free parameters (the values of which were found not to influence results): $\nu=\mu = \SI{1e-8}{}$ (i.e. a magnetic Prandtl number of one), $\nu_L=\mu_L = \SI{1e1}{}$, $k_0=1$, and $N=76$. 
The modes $u_{n}^{\Sk}$ and $B_{n}^{\Sk}$ were initialized in a K41-scaling state $\sim k_n^{-2/3}$ (the shell model equivalent of $\sim k^{-5/3}$) with zero helicity of any kind, and stepped forward using a fourth-order Runge--Kutta integration scheme with a time-step of $\Delta t = \SI{5e-7}{}$. 
The injection rates $\meanenergyu=\meanenergyB=1/2$, and $\meanhelicityu,\meanhelicityB,\meanhelicitycro=0$ (i.e. no forcing handedness, elaborated on in the discussion) were evenly applied over $p$ shells, starting from shell $n_f = 30$. Thus, in the $p=3$ configuration, shells \#31 and \#32 were also forced.
Note that while dynamo studies typically inject only kinetic energy, both kinetic and magnetic energy were injected in the present simulations. 
Although no difference in the energy flux partitioning between MTI components was found for homochiral MTIs $i=1,2,3$, the magnetic field collapsed for $i=4$ unless magnetic energy was injected.

The simulated partial energy fluxes within the four homochiral configurations $i=1,2,3,4$ are shown in figures \ref{fig:MHDcoupled}a, \ref{fig:MHDcoupled}b, \ref{fig:MHDcoupled}c, and \ref{fig:MHDcoupled}d, respectively, calculated using \eqnref{eqn:SMenergyfluxA}--\eqnref{eqn:SMenergyfluxD}.
The energy flux partitionings are shown for both triad shapes, labeled in the legends by the corresponding $\alphaBuB$ exponents. 
Gray/black lines represent components without pseudo-invariants, whereas components marked by coloured lines posses pseudo-invariants: 
blue colors indicate components for which the pseudo-invariant exponents are negative ($\alpha<0$, hypothesised forward contribution), whereas red/orange colors mark components conserving enstrophy-like quantities with positive exponents ($\alpha>0$, hypothesised inverse contribution).

Figures \ref{fig:MHDcoupled}a and \ref{fig:MHDcoupled}b show the partitionings within homochiral MTIs $i=1$ and $i=2$ conform with the pseudo-invariant predictions:
all MTI components contribute to a forward transfer, whereas $\GBuBl{2}$ (conserving an enstrophy-like quantity, $\alphaBuB>0$) contributes to an inverse transfer.

Figure \ref{fig:MHDcoupled}c shows the partitioning within MTI $i=3$ almost behaving as expected: all components contribute to a forward transfer of energy, except for $\GBuBl{3}$ and $\GBBum{3}$ interactions which have their behaviours reversed --- i.e. components conserving enstrophy-like quantities are found to contribute to a forward energy transfer and vice-versa.

Figure \ref{fig:MHDcoupled}d shows the partitioning within MTI $i=4$ also conforms with predictions: all components contribute to a forward transfer of energy, whereas $\GBBum{4}$ (conserving an enstrophy-like quantity, $\alphaBBu>0$) contributes inversely.
For the simulated triad shape corresponding to $\alphaBuB=-2.0$, however, the $\GBuBl{4}$ component carries a small inverse contribution in spite of not conserving any enstrophy-like quantity.

\added{
Finally, in the nonlinear regime considered here (non-negligible back-reaction of the Lorentz force on the flow), we note that the partial fluxes are more or less constant, suggesting that relatively local interactions in wave space may facilitate both up- and downscale embedded energy cascades.
}

\section{Discussion}
\label{sec:discussion}
\added{
The simulated energy flux partitionings between MTI components largely (but not fully) support the proposed influence of the new pseudo-invariants, which allow the energy flux partitionings to be understood in terms of conserved quantities in analogy to enstrophy in 2D HD turbulence. 
In this section, we further discuss and investigate in detail the relevance of our results to larger triad networks (section \ref{sec:largenetworks}), the effect of coupling different MTI components and the mismatch between some of the predictions and simulations (section \ref{sec:couplingeffect}), the possibility of instigating an energy transfer reversal by biasing triad interactions (section \ref{sec:transitionclassifications}), the scale locality of the triad interactions which may facilitate large-scale $\alpha$-dynamo action and the inverse transfer of magnetic helicity (section \ref{sec:scalelocality}), and caveats about the robustness of our results to various forcing scenarios (section \ref{sec:universality}).
}

\added{
\subsection{Relevance to large triad networks}
\label{sec:largenetworks}
The relevance of the new partial invariants to more realistic triad configurations consisting of large MTI networks may be surmised by comparing with recent direct numerical simulations (DNSs) of \eqnref{eqn:MHDrealspace} or \eqnref{eqn:MHDa}--\eqnref{eqn:MHDb}. 
The DNS by \citet{bib:linkmann2017triad} considered a nonhelical and a positive helical forcing of the velocity and magnetic fields, respectively, thus introducing a dominant sign of magnetic helicity.
In this case, both forward and inverse (bidirectional) magnetic helicity cascades were identified, and the inverse component was attributed to (dominant) contributions from $\GBBum{3}$ and $\GBBum{4}$ triads which involve interactions between magnetic components of same helical sign (Figure \ref{fig:triadtypes_MHD}c). 
However, in the steady-state regime, low-wavenumber modes were found to develop negative helicity spectra despite a mean positive value due to pumping (and distributed by a cascade process).
Because the forcing scenario led to a positive kinetic helicity spectrum (associated with the action of the Lorentz force) the negative helicity spectrum at large scales was argued to be the result of large-scale dynamo action facilitated by $\GBuBl{2}$ triads, since only the action of such triads was consistent with the simulated helical signature ($\alpha$-dynamo signature).

\citet{bib:linkmann2017effects} also considered strongly magnetized flows by using a helical magnetic forcing in a DNS of \eqnref{eqn:MHDa}--\eqnref{eqn:MHDb}.
They noted that if the triads that facilitate the inverse transfer of magnetic helicity indeed are the dominant ones from a linear stability analysis, then the transport efficiency depends on the degree to which small-scale magnetic and kinetic helicity is of the same sign, which they confirmed numerically. 
Thus, dominant triad interactions which have large growth rates of perturbed modes from a stability analysis, such as $\GBBum{4}$, might play an important role in the transfer of magnetic helicity.
Their instability analysis also predicted that triads with helical signatures reminiscent of an $\alpha$-type dynamo should be dominant due to large growth rates, which too was confirmed numerically in dynamo experiments using a purely mechanical forcing. Therefore, by similar arguments, $\GBuBl{2}$ triads might play an important role in large triad networks by facilitating large-scale $\alpha$-dynamo action.

In short, the pseudo-invariants may, arguably, have relevance for larger triad networks in the sense that the triads we predict to facilitate inverse energy transfers have previously been suggested important in DNS studies (and triad stability analyses) for large-scale magnetic structure formation by large-scale dynamo action and the inverse transfer of magnetic helicity.
}

\begin{figure*}[!t]
\includegraphics[width=1\textwidth]{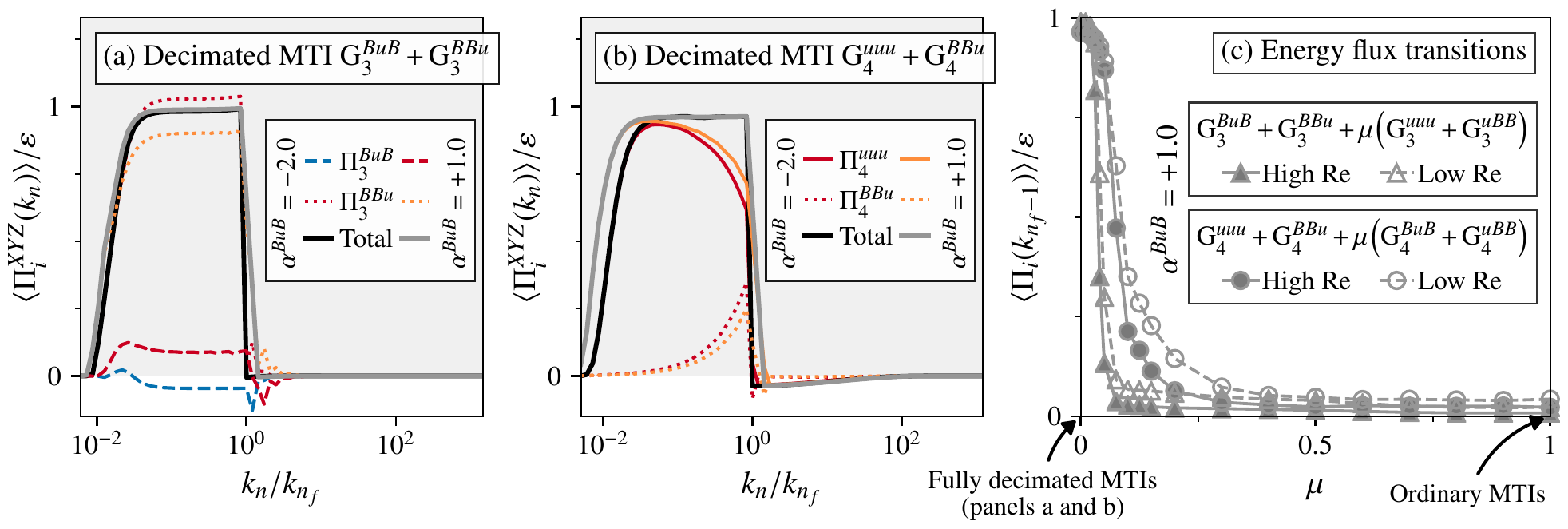} 
\caption{
Panels \textbf{a} and \textbf{b}: simulated partial energy fluxes associated with each MTI-component of the decimated MTIs $\GBuBl{3}+\GBBum{3}$ and $\Guuui{4}+\GBBum{4}$.
Negative (positive) values correspond to forward/downscale (inverse/upscale) fluxes. 
Blue lines denote hypothesised forward partial fluxes according to the pseudo-invariant predictions, while red/orange lines denote hypothesised inverse partial fluxes (interactions conserving an enstrophy-like quantity). 
\added{
Panel \textbf{c}: inverse energy flux as a function of the decimation (control) parameter, $\mu$, for the two MTIs 
$\GBuBl{3}+\GBBum{3} + \mu(\Guuui{3}+\GuBBj{3})$ and 
$\Guuui{4}+\GBBum{4} + \mu(\GBuBl{4}+\GuBBj{4})$, defined as the total flux magnitude immediately above the forcing scale, $k_{n_f-1}$, normalized by the pumping rate, $\meanenergy$. 
Note that the MTIs are fully decimated for $\mu=0$ (partitionings shown in panels \textbf{a} and \textbf{b}), while for $\mu=1$ the ordinary MTIs are recovered (partitionings shown in Figure \ref{fig:MHDcoupled}c and \ref{fig:MHDcoupled}d).
Filled markers denote high Reynolds number simulations (canonical model configuration used throughout), whereas empty markers denote low Reynolds number simulations (large-scale viscosities increased by a factor of 10).
}
\label{fig:MHDdecoupled}}
\end{figure*}

\subsection{Effect of coupling MTI components}
\label{sec:couplingeffect}
The homochiral MTIs of sign groups $i=3$ and $i=4$ show discrepancies between simulated partial fluxes and predictions based on the pseudo-invariants (section \ref{sec:MHDresults}).
\added{Interestingly, for the homochiral MTI $i=3$, these discrepancies correspond to triad interactions with low growth rates of the perturbed modes in a linear stability analysis \citep{bib:linkmann2017triad} (to be understood why).}
In the case of the shell model, we find that coupling-induced effects, caused by the coupling of MTI components, might play an important role by altering the component-wise behaviours (partial fluxes) compared to predictions. 
This is exemplified by considering additional decimated simulations in which only $\GBuB$ and $\GBBu$ are coupled, i.e. decomposing the MTIs by disregarding $\Guuu$ and $\GuBB$ interactions. 
Using the same model configuration as above, Figure \ref{fig:MHDdecoupled}a shows the resulting energy flux partitioning between the two components $\GBuBl{3}$ and $\GBBum{3}$ in the decimated MTI $\GBuBl{3}+\GBBum{3}$. 
For all four decimated MTIs $\GBuBl{i}+\GBBum{i}$ ($i=1,2,3,4$), the component-wise partial fluxes are found to conform with the pseudo-invariant predictions ($i=1,2,4$ not shown), suggesting the discrepancies between the full and decimated MTIs might be due to coupling-induced effects.
Whether this is a model artefact or a property also shared by large triad networks \eqnref{eqn:MHDa}--\eqnref{eqn:MHDb} is not clear.

\added{
\subsection{Classification of energy flux transitions}
\label{sec:transitionclassifications}
Efforts have previously been made to classify how HD and MHD systems may transition from a state in which energy is transferred (e.g. by a cascade process) primarily downscale to primarily upscale as a function of external mechanisms such as rotation, confinement, stratification, and background magnetic fields, and by which triad interactions energy is channelled upscale [see \citet{alexakis2018cascades} for a review].
For example, by numerically varying the thickness of a nonconducting fluid layer, a possible critical thickness has been reported at which a continuous but nonsmooth (second order) transition occours from a forward to inverse energy cascade \citep{bib:benavides2017critical}.
By considering a nonconducting fluid in a rotating reference frame, \citet{buzzicotti2018energy} recently numerically explored the transition from a forward cascade to a split forward--inverse cascade as a function of rotation rate, finding that $\Guuui{4}$ HD triad interactions---which conserve enstrophy-like quantities by conserving both signs of kinetic helicity separately---might predominantly be responsible for channelling energy upscale under rapid rotation.

Inverse energy cascades in 3D MHD turbulence, such as found in the presence of strong background (guiding) magnetic fields both numerically \citep{alexakis2011two} and experimentally \citep{baker2018inverse}, might also be related to subsets of helical triad interactions conserving enstrophy-like quantities.
This is indeed possible in principle as shown in Figure \ref{fig:MHDdecoupled}a, which is a robust result for other combinations of enstrophy-like conserving interactions, such as shown in Figure \ref{fig:MHDdecoupled}b for the decimated MTI $\Guuui{4}+\GBBum{4}$.

In an attempt to explore whether favouring interactions conserving enstrophy-like quantities can induce a predominantly inverse transfer of energy in MHD turbulence, we conducted a set of triad-biasing (decimation) experiments inspired by \citet{bib:sahoo2015roleof}, \citet{bib:sahoo2015disentangling}, and \citet{bib:sahoo2017helicity,bib:sahoo2017discontinuous}. Specifically, we considered the decimation of the two homochiral MTIs $i=3$ and $i=4$ using a control parameter, $\mu$, such that only enstrophy-like conserving interactions are considered for $\mu=0$ (corresponding partitionings shown in Figures \ref{fig:MHDdecoupled}a and \ref{fig:MHDdecoupled}b) while all interactions are considered without bias for $\mu=1$ (corresponding partitionings shown in Figures \ref{fig:MHDcoupled}c and \ref{fig:MHDcoupled}d), that is
\begin{align}
\GBuBl{3}+\GBBum{3} &+ \mu\left(\Guuui{3}+\GuBBj{3}\right) \label{eqn:demarcMTIa}
\\
\Guuui{4}+\GBBum{4} &+ \mu\left(\GBuBl{4}+\GuBBj{4}\right) \label{eqn:demarcMTIb}
.
\end{align} 
Figure \ref{fig:MHDdecoupled}c shows the resulting total inverse energy flux immediately above the forcing scale, $k_{n_f-1}$, as a function of $\mu$ for the two decimated MTIs \eqnref{eqn:demarcMTIa} (triangles) and \eqnref{eqn:demarcMTIb} (circles) with triad shapes corresponding to $\alphaBuB=+1.0$. 
Following the terminology in \citet{alexakis2018cascades}, we find that the cascade transitions are continuous and smooth from strictly forward to strictly inverse, but might tend toward discontinuous (first-order) transitions in the limit of long inertial--inductive ranges [difference in Figure \ref{fig:MHDdecoupled}c between full markers (high Reynolds number, canonical model configuration) and empty markers (low Reynolds number, large-scale viscosities increased by a factor of 10)].

\citet{bib:sahoo2017discontinuous} recently studied the decimated HD system $\mu(\Guuui{1}+\Guuui{2}+\Guuui{3})+\Guuui{4}$, finding also that the transition toward an inverse energy cascade as a function of $\mu$ becomes a discontinuous jump in the limit of large Reynolds number. 
In addition, they found that the transition appears to be quasi singular in the sense that the critical control parameter value is close to $\mu=1$.
Here, we find that this result also carries over to MHD systems, suggesting that enstrophy-like conserving interactions are comparatively less efficient at transferring energy, and thus only a small percentage of forward-transferring interactions are necessary to maintain a predominantly forward transfer of energy.
}

\added{
\subsection{Scale locality of interactions}
\label{sec:scalelocality}
In HD turbulence, the relative importance of local versus nonlocal triad interactions has been vigorously investigated both theoretically and numerically.
Although numerical simulations have demonstrated that nonlocal interactions involving large-scale flow components can influence the energy cascade \citep{domaradzki1990local,alexakis2005imprint,mininni2006large}, thus challenging to the Kolmogorov picture of a local energy cascade, other numerical studies considering high Reynolds number flows \citep{domaradzki2007analysis,mininni2008nonlocal,domaradzki2009locality} and theoretical considerations \citep{eyink2009localness,aluie2009localness} suggest that nonlocal contributions to the energy cascade are significantly smaller than local contributions. 
Thus, while individual nonlocal interactions (having one leg at the energy-containing scales) may contribute more to the energy flux than individual local interactions, the act of summing over all triad contributions is dominated by local interactions \citep{domaradzki2007analysis}.

In the MHD case, the picture is less clear, but there is a growing consensus---at least for intermediate Reynolds numbers---that MHD turbulence is less local than HD turbulence \citep{mininni2011scale}.
Depending on the flow configuration (forced, decaying, presence of external fields), nonlocal transfers have been found in MHD turbulence which involve interactions between velocity and magnetic fields \citep{bib:brandenburg2005astrophysical,verma2005energy,alexakis2005shell,carati2006energy}, or related to the transfer of magnetic helicity \citep{alexakis2006inverse,aluie2010scale}, while velocity--velocity and magnetic--magnetic field interactions are mostly local \citep{mininni2011scale}.

The possible relevance of the pseudo-invariants for transfer processes in MHD turbulence depends, therefore, on the degree to which the aggregate of triads conserving them are relevant for the velocity and magnetic field evolutions. 
Since the HD interactions $\Guuui{2}$ conserve enstrophy-like invariants only in the nonlocal limit, their relevance is arguably small insofar as HD flows are dominated by local interactions.
We note, however, that the flow configuration (e.g. aspect ratio, rotation) might influence which triads are predominantly responsible for channelling energy \citep{alexakis2018cascades}, suggesting care must be taken when determining the relative importance of triads. 
Conversely, in the case of MHD turbulence, the enstrophy-like conservations are found in both the local and nonlocal (acute triad) parts of interaction space and for both pseudo-invariants (red colors in Figure \ref{fig:triadtypes_MHD}a and \ref{fig:triadtypes_MHD}b).

Whether the pseudo-invariants are relevant only for constraining cascade contributions to the total energy flux, or more generally applicable to spectral transfers (process not associated with a constant flux), is an open question. 
We note, however, that the enstrophy-like conservation by $\GBBum{3}$ and $\GBBum{4}$ triads, which have magnetic components of like-signed helicity as the two smallest triad legs (largest scales), supports the possibility of an inverse magnetic energy transfer consistent with an inverse magnetic helicity cascade, as pointed out by \citet{bib:linkmann2016helical} and \citet{bib:linkmann2017triad}.
The inter-scale transfer of magnetic helicity is complex, with concurrent forward and inverse contributions \citep{mininni2011scale,bib:linkmann2017triad}. 
On this matter, it is not clear whether the inverse component (and accompanied energy transfers) is dominated by local or nonlocal transfers \citep{alexakis2006inverse,aluie2010scale,mininni2011scale,muller2012inverse,linkmann2016large,bib:linkmann2017effects}. 
Unfortunately, this work provides no further clarity since the dependence of the pseudo-invariant on triad geometry is consistent with both local and nonlocal inverse magnetic helicity transfers being possible. We do, however, find that local transfers can take the form of a (constant flux) cascade process. 

If the pseudo-invariants are indeed generally applicable to spectral transfers, the agreement is intriguing between the (triadic) large-scale dynamo action suggested to be facilitated by $\GBuBl{2}$ and $\GBuBl{3}$ triads in the nonlocal limit ($\VKLEN\ll \VPLEN\approx\VQLEN$) \citep{bib:linkmann2016helical,bib:linkmann2017effects} and the enstrophy-like conservations. 
Of the two, $\GBuBl{2}$ interactions have the same helical signature as the $\alpha$-effect \citep{bib:linkmann2016helical}, and have stability properties suggesting a dominant role over $\GBuBl{3}$ interactions.
Although care must be taken when extrapolating the behaviour of single triads to a large system of triads \citep{moffatt2014note}, we note that the enstrophy-like conservation by $\GBuBl{2}$ (and $\GBuBl{3}$) (Figure \ref{fig:triadtypes_MHD}a) suggests that inverse transfers from such triads might be dominated by nonlocal interactions compared to $\GBBum{3}$ and $\GBBum{4}$ triads (Figure \ref{fig:triadtypes_MHD}b); that is, the inverse transfer of magnetic helicity is more local than large-scale $\alpha$-dynamo action.
}

\subsection{Universality of energy flux partitioning}
\label{sec:universality}
\begin{figure}[!t]
\includegraphics[width=1\columnwidth]{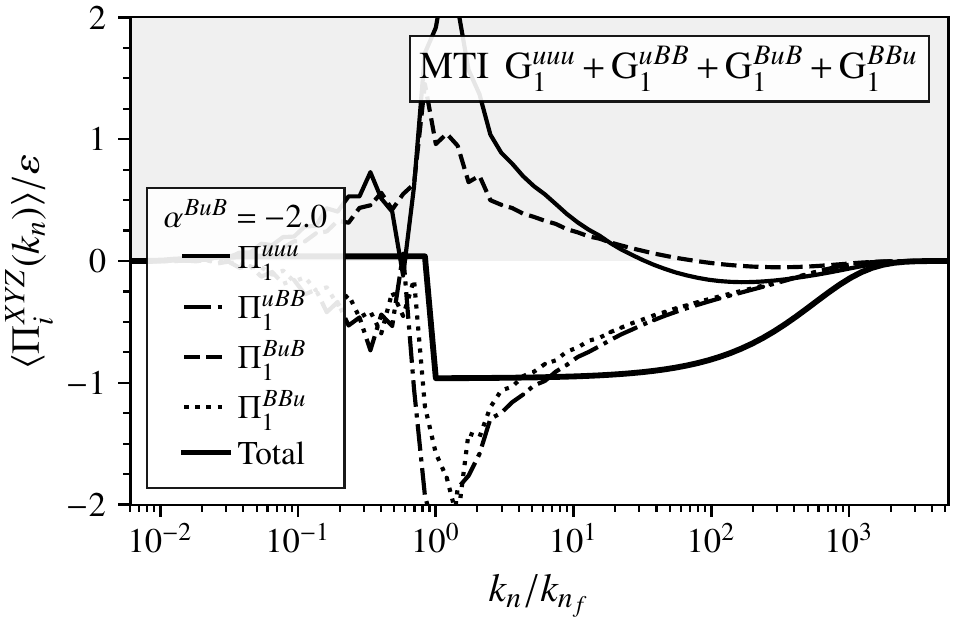} 
\caption{
Simulated energy flux partitioning between the four MTI components within the homochiral MTI $\Guuui{1}+\GuBBj{1}+\GBuBl{1}+\GBBum{1}$ for a nonzero cross-helicity pumping of $\meanhelicitycrou=0.25$.
Negative (positive) values correspond to forward/downscale (inverse/upscale) fluxes. 
\label{fig:MHDdecoupled_frcdep}}
\end{figure}

The partitioning of the energy cascade between symmetrized helical triad interactions was recently investigated by \citet{bib:alexakis2016helically} in a direct numerical simulation of the spectral--helical Navier--Stokes equation.
Like here, it was found that the energy cascade partitions itself into approximately constant partial-flux components within the inertial range, an intriguing result given that only the total energy flux is required to be constant.
Moreover, \citet{bib:alexakis2016helically} found that the partitioning was unaffected by the pumping of kinetic helicity, suggesting the partitioning might be universal (among a given set of resolved triads).
If so, the physical explanation for the directionalities of the partial fluxes in terms of pseudo-invariants might be applicable in a general, forcing-independent sense.

Inspired by this result, we conducted additional simulations where the four homochiral MTIs  were forced with kinetic helicity ($\meanhelicityu$), magnetic helicity ($\meanhelicityB$), and cross-helicity ($\meanhelicitycro$) besides energy, in contrast to the nonhelical forcing used above ($\meanhelicityu,\meanhelicityB,\meanhelicitycro=0$).
While the partitionings were found to be independent of $\meanhelicityu$ and $\meanhelicityB$, suggesting the partitionings might be universal, the same was not found for a nonzero injection of cross-helicity ($\meanhelicitycro\neq0$).
This is exemplified in Figure \ref{fig:MHDdecoupled_frcdep} for the homochiral MTI $i=1$ (similar results were found for $i=2,3,4$), demonstrating the divergent, non-constant partial fluxes which develop in the case of $\meanhelicitycrou=0.25$. 
In addition, when pumping cross-helicity, the simulated directionalities of the partial fluxes generally do not agree with the pseudo-invariant predictions: in Figure \ref{fig:MHDdecoupled_frcdep}, all components are predicted to contribute to a forward energy transfer since no enstrophy-like quantities are conserved.

Because the evolution of magnetic modes depends on the alignment between velocity and magnetic modes (cross-helicity) according to \eqnref{eqn:MHDb}, it is not entirely surprising that injecting cross-helicity can affect the detailed partitioning.
Note that this result is in agreement with a linear stability analysis which also predicts that triad leg (modal) stabilities depend on velocity--magnetic alignments \citep{bib:linkmann2016helical}.

The present analysis considered only a magnetic Prandtl number of one. While investigating large and small magnetic Prandtl number flows is out of scope of the present work, studies such as \citet{bib:verma2016dynamos} indicate that under some forcing conditions the partitioning might be unaffected, although helically decomposed dynamics were not considered in that case.

\section{Conclusion}
In conclusion, we identified new, partially conserved quantities among the elementary three-wave (triad) interactions in spectral--helically decomposed ideal magnetohydrodynamical (MHD) turbulence.  
Because the new quantities conserved by a subset of triad interactions are enstrophy-like, we conjecture that such interactions might contribute to embedded, inverse energy transfers developing in three-dimensional MHD turbulence in analogy to enstrophy-conserving triad interactions in two-dimensional hydrodynamical (HD) turbulence.

\added{
Among the enstrophy-like conserving interactions are the helical interactions recently identified with facilitating large-scale $\alpha$-dynamo action ($\GBuBl{2}$) and the inverse (upscale) transfer of magnetic helicity ($\GBBum{4}$) (Figure \ref{fig:triadtypes_MHD}c). 
The partial invariants (pseudo-invariants) might therefore play an important role in the understanding of which physical mechanisms are behind the transfer processes leading to large-scale magnetic structure formation, and therefore of astrophysical interest.
Importantly, the conservation of the pseudo-invariants depend on the scale locality of the triad interactions (Figure \ref{fig:triadtypes_MHD}a and \ref{fig:triadtypes_MHD}b). Based on this, we conjecture that large-scale dynamo action with an $\alpha$-type helical signature (large- and small-scale magnetic field components having opposite signs of helicity) is a transfer process that is more nonlocal in wave space than the inverse transfer of magnetic helicity, insofar as they are facilitated by $\GBuBl{2}$ and $\GBBum{4}$ triads, respectively. 
}

In order to test the predictions based on the new pseudo-invariants, we introduced a helically decomposed reduced wave-space model (shell model). 
By conducting numerical simulations of the four kinds of homochiral minimal sets of triad interactions (MTIs) (minimal in the sense of being required to conserve the ideal MHD invariants), we demonstrated the usefulness of the partial invariants for understanding the resulting embedded energy flux contributions (partial fluxes) from the triadic components constituting an MTI.
\added{
Inspired by recent HD studies showing that the relative importance of different triad interactions might depend on flow configuration (e.g. rotation or confinement), we additionally demonstrated the possibility of a strictly inverse transfers of energy developing if enstrophy-like conserving interactions are favoured, finding that only a small percentage of triads not conserving enstrophy-like quantities are necessary to maintain a dominant downscale transfer of energy (magnetic+kinetic).
}

While for simplicity this study concerned itself with the case of a magnetic Prandtl number of one, we find that the embedded partial fluxes are generally constant over inertial--inductive ranges, indicating that forward and inverse energy cascades might generally exist embedded in MHD turbulence.
Importantly, however, we note that the injection of cross-helicity by the forcing mechanism, \replaced{}{and the effect of coupling certain types of MTI components (triad interactions), demonstrate cases where the directions of the simulated partial fluxes do not agree with the pseudo-invariant predictions. Whether this is a model artefact or a property shared by more comprehensive (and thus more realistic) large triad networks as represented by the spectral--helically decomposed MHD equations \eqnref{eqn:MHDa}--\eqnref{eqn:MHDb}, is not clear.}

\acknowledgments

\added{
The authors wish to thank an anonymous referee for useful comments and references that led to an appreciably improved manuscript which, importantly, includes the connection between the (triadic) $\alpha$-dynamo and the $\GBuBl{2}$ pseudo-invariant, and Luca Biferale for comments and suggestions leading to section \ref{sec:transitionclassifications}.
}

\bibliography{cascadepartitioning}

\begin{thebibliography}{}
\expandafter\ifx\csname natexlab\endcsname\relax\def\natexlab#1{#1}\fi
\providecommand{\url}[1]{\href{#1}{#1}}
\providecommand{\dodoi}[1]{doi:~\href{http://doi.org/#1}{\nolinkurl{#1}}}
\providecommand{\doeprint}[1]{\href{http://ascl.net/#1}{\nolinkurl{http://ascl.net/#1}}}
\providecommand{\doarXiv}[1]{\href{https://arxiv.org/abs/#1}{\nolinkurl{https://arxiv.org/abs/#1}}}

\bibitem[{Alexakis(2011)}]{alexakis2011two}
Alexakis, A. 2011, Physical Review E, 84, 056330

\bibitem[{Alexakis(2017)}]{bib:alexakis2016helically}
---. 2017, Journal of Fluid Mechanics, 812, 752

\bibitem[{Alexakis \& Biferale(2018)}]{alexakis2018cascades}
Alexakis, A., \& Biferale, L. 2018, Physics Reports

\bibitem[{Alexakis {et~al.}(2005{\natexlab{a}})Alexakis, Mininni, \&
  Pouquet}]{alexakis2005imprint}
Alexakis, A., Mininni, P., \& Pouquet, A. 2005{\natexlab{a}}, Physical review
  letters, 95, 264503

\bibitem[{Alexakis {et~al.}(2005{\natexlab{b}})Alexakis, Mininni, \&
  Pouquet}]{alexakis2005shell}
Alexakis, A., Mininni, P.~D., \& Pouquet, A. 2005{\natexlab{b}}, Physical
  Review E, 72, 046301

\bibitem[{Alexakis {et~al.}(2006)Alexakis, Mininni, \&
  Pouquet}]{alexakis2006inverse}
---. 2006, The Astrophysical Journal, 640, 335

\bibitem[{Aluie(2017)}]{aluie2017coarse}
Aluie, H. 2017, New Journal of Physics, 19, 025008

\bibitem[{Aluie \& Eyink(2009)}]{aluie2009localness}
Aluie, H., \& Eyink, G.~L. 2009, Physics of Fluids, 21, 115108

\bibitem[{Aluie \& Eyink(2010)}]{aluie2010scale}
---. 2010, Physical review letters, 104, 081101

\bibitem[{Baker {et~al.}(2018)Baker, Poth{\'e}rat, Davoust, \&
  Debray}]{baker2018inverse}
Baker, N.~T., Poth{\'e}rat, A., Davoust, L., \& Debray, F. 2018, Physical
  review letters, 120, 224502

\bibitem[{Benavides \& Alexakis(2017)}]{bib:benavides2017critical}
Benavides, S.~J., \& Alexakis, A. 2017, Journal of Fluid Mechanics, 822, 364

\bibitem[{Benzi {et~al.}(1996)Benzi, Biferale, Kerr, \&
  Trovatore}]{bib:benzi1996helical}
Benzi, R., Biferale, L., Kerr, R., \& Trovatore, E. 1996, Physical Review E,
  53, 3541

\bibitem[{Bian \& Aluie(2019)}]{bian2019decoupled}
Bian, X., \& Aluie, H. 2019, Physical review letters, 122, 135101

\bibitem[{Biferale {et~al.}(2012)Biferale, Musacchio, \&
  Toschi}]{bib:biferale2012inverse}
Biferale, L., Musacchio, S., \& Toschi, F. 2012, Physical review letters, 108,
  164501

\bibitem[{Biskamp(1993)}]{bib:biskamp1993nonlinear}
Biskamp, D. 1993, Nonlinear magnetohydrodynamics, volume 1 of Cambridge
  Monographs on Plasma Physics,  Cambridge University Press, Cambridge

\bibitem[{Blackman(2016)}]{blackman2016magnetic}
Blackman, E.~G. 2016, in Multi-scale Structure Formation and Dynamics in Cosmic
  Plasmas (Springer), 59--91

\bibitem[{Boffetta \& Musacchio(2010)}]{bib:boffetta2010evidence}
Boffetta, G., \& Musacchio, S. 2010, Physical Review E, 82, 016307

\bibitem[{Brandenburg(2001)}]{brandenburg2001inverse}
Brandenburg, A. 2001, The Astrophysical Journal, 550, 824

\bibitem[{Brandenburg {et~al.}(2015)Brandenburg, Kahniashvili, \&
  Tevzadze}]{brandenburg2015nonhelical}
Brandenburg, A., Kahniashvili, T., \& Tevzadze, A.~G. 2015, Physical review
  letters, 114, 075001

\bibitem[{Brandenburg \& Subramanian(2005)}]{bib:brandenburg2005astrophysical}
Brandenburg, A., \& Subramanian, K. 2005, Physics Reports, 417, 1

\bibitem[{Brissaud {et~al.}(1973)Brissaud, Frisch, Leorat, Lesieur, \&
  Mazure}]{bib:brissaud1973helicity}
Brissaud, A., Frisch, U., Leorat, J., Lesieur, M., \& Mazure, A. 1973, Physics
  of Fluids (1958-1988), 16, 1366

\bibitem[{Brun \& Browning(2017)}]{brun2017magnetism}
Brun, A.~S., \& Browning, M.~K. 2017, Living Reviews in Solar Physics, 14, 4

\bibitem[{Buzzicotti {et~al.}(2018)Buzzicotti, Aluie, Biferale, \&
  Linkmann}]{buzzicotti2018energy}
Buzzicotti, M., Aluie, H., Biferale, L., \& Linkmann, M. 2018, Physical Review
  Fluids, 3, 034802

\bibitem[{Carati {et~al.}(2006)Carati, Debliquy, Knaepen, Teaca, \&
  Verma}]{carati2006energy}
Carati, D., Debliquy, O., Knaepen, B., Teaca, B., \& Verma, M. 2006, Journal of
  Turbulence, N51

\bibitem[{Childress \& Gilbert(1995)}]{childress1995stretch}
Childress, S., \& Gilbert, A.~D. 1995, Stretch, twist, fold: the fast dynamo,
  Vol.~37 (Springer Science \& Business Media)

\bibitem[{Constantin \& Majda(1988)}]{bib:constantin1988beltrami}
Constantin, P., \& Majda, A. 1988, Communications in mathematical physics, 115,
  435

\bibitem[{De~Pietro {et~al.}(2015)De~Pietro, Biferale, \&
  Mailybaev}]{bib:DePietro2015}
De~Pietro, M., Biferale, L., \& Mailybaev, A.~A. 2015, Phys. Rev. E, 92,
  043021, \dodoi{10.1103/PhysRevE.92.043021}

\bibitem[{De~Pietro {et~al.}(2017)De~Pietro, Mailybaev, \&
  Biferale}]{bib:DePietro2016chaotic}
De~Pietro, M., Mailybaev, A.~A., \& Biferale, L. 2017, Physical Review Fluids,
  2, 034606

\bibitem[{D{\'e}moulin(2007)}]{demoulin2007recent}
D{\'e}moulin, P. 2007, Advances in Space Research, 39, 1674

\bibitem[{Ditlevsen \&
  Giuliani(2001{\natexlab{a}})}]{bib:ditlevsen2001dissipation}
Ditlevsen, P.~D., \& Giuliani, P. 2001{\natexlab{a}}, Physics of Fluids
  (1994-present), 13, 3508

\bibitem[{Ditlevsen \&
  Giuliani(2001{\natexlab{b}})}]{bib:ditlevsen2001cascades}
---. 2001{\natexlab{b}}, Physical Review E, 63, 036304

\bibitem[{Domaradzki \& Carati(2007)}]{domaradzki2007analysis}
Domaradzki, J.~A., \& Carati, D. 2007, Physics of fluids, 19, 085112

\bibitem[{Domaradzki \& Rogallo(1990)}]{domaradzki1990local}
Domaradzki, J.~A., \& Rogallo, R.~S. 1990, Physics of Fluids A: Fluid Dynamics,
  2, 413

\bibitem[{Domaradzki {et~al.}(2009)Domaradzki, Teaca, \&
  Carati}]{domaradzki2009locality}
Domaradzki, J.~A., Teaca, B., \& Carati, D. 2009, Physics of fluids, 21, 025106

\bibitem[{Eyink \& Aluie(2009)}]{eyink2009localness}
Eyink, G.~L., \& Aluie, H. 2009, Physics of Fluids, 21, 115107

\bibitem[{Frisch {et~al.}(1975)Frisch, Pouquet, L{\'e}orat, \&
  Mazure}]{bib:frisch1975possibility}
Frisch, U., Pouquet, A., L{\'e}orat, J., \& Mazure, A. 1975, Journal of Fluid
  Mechanics, 68, 769

\bibitem[{G{\"u}rcan(2017)}]{gurcan2017nested}
G{\"u}rcan, {\"O}. 2017, Physical Review E, 95, 063102

\bibitem[{G{\"u}rcan(2018)}]{gurcan2018nested}
---. 2018, Physical Review E, 97, 063111

\bibitem[{Hood \& Hughes(2011)}]{hood2011solar}
Hood, A.~W., \& Hughes, D.~W. 2011, Physics of the Earth and Planetary
  Interiors, 187, 78

\bibitem[{Howes \& Quataert(2010)}]{howes2010interpretation}
Howes, G.~G., \& Quataert, E. 2010, The Astrophysical Journal Letters, 709, L49

\bibitem[{Kraichnan(1967)}]{bib:kraichnan1967inertial}
Kraichnan, R.~H. 1967, Inertial ranges in two-dimensional turbulence, Tech.
  rep., DTIC Document

\bibitem[{Kraichnan(1973)}]{bib:kraichnan1973helical}
---. 1973, Journal of Fluid Mechanics, 59, 745

\bibitem[{Krause \& R{\"a}dler(1980)}]{krause1980mean}
Krause, F., \& R{\"a}dler, K. 1980, Akademieverlag, Berlin

\bibitem[{Lessinnes {et~al.}(2009)Lessinnes, Plunian, \&
  Carati}]{bib:lessinnes2009helical}
Lessinnes, T., Plunian, F., \& Carati, D. 2009, Theoretical and Computational
  Fluid Dynamics, 23, 439

\bibitem[{Lessinnes {et~al.}(2011)Lessinnes, Plunian, Stepanov, \&
  Carati}]{bib:lessinnes2011dissipation}
Lessinnes, T., Plunian, F., Stepanov, R., \& Carati, D. 2011, Physics of Fluids
  (1994-present), 23, 035108

\bibitem[{Linkmann {et~al.}(2016)Linkmann, Berera, McKay, \&
  J{\"a}ger}]{bib:linkmann2016helical}
Linkmann, M., Berera, A., McKay, M., \& J{\"a}ger, J. 2016, Journal of Fluid
  Mechanics, 791, 61

\bibitem[{Linkmann \& Dallas(2016)}]{linkmann2016large}
Linkmann, M., \& Dallas, V. 2016, Physical Review E, 94, 053209

\bibitem[{Linkmann \& Dallas(2017)}]{bib:linkmann2017triad}
---. 2017, Physical Review Fluids, 2, 054605

\bibitem[{Linkmann {et~al.}(2017)Linkmann, Sahoo, McKay, Berera, \&
  Biferale}]{bib:linkmann2017effects}
Linkmann, M., Sahoo, G., McKay, M., Berera, A., \& Biferale, L. 2017, The
  Astrophysical Journal, 836, 26

\bibitem[{Malapaka \& M{\"u}ller(2013{\natexlab{a}})}]{malapaka2013modeling}
Malapaka, S.~K., \& M{\"u}ller, W.-C. 2013{\natexlab{a}}, The Astrophysical
  Journal, 774, 84

\bibitem[{Malapaka \& M{\"u}ller(2013{\natexlab{b}})}]{malapaka2013large}
---. 2013{\natexlab{b}}, The Astrophysical Journal, 778, 21

\bibitem[{Mininni {et~al.}(2006)Mininni, Alexakis, \&
  Pouquet}]{mininni2006large}
Mininni, P., Alexakis, A., \& Pouquet, A. 2006, Physical Review E, 74, 016303

\bibitem[{Mininni(2011)}]{mininni2011scale}
Mininni, P.~D. 2011, Annual Review of Fluid Mechanics, 43, 377

\bibitem[{Mininni {et~al.}(2008)Mininni, Alexakis, \&
  Pouquet}]{mininni2008nonlocal}
Mininni, P.~D., Alexakis, A., \& Pouquet, A. 2008, Physical review E, 77,
  036306

\bibitem[{Mininni {et~al.}(2009)Mininni, Alexakis, \&
  Pouquet}]{bib:mininni2009scale}
---. 2009, Physics of Fluids (1994-present), 21, 015108

\bibitem[{Mininni \& Pouquet(2013)}]{bib:mininni2013inverse}
Mininni, P.~D., \& Pouquet, A. 2013, Physical Review E, 87, 033002

\bibitem[{Moffatt(1989)}]{moffatt1989stretch}
Moffatt, H. 1989, Nature, 341, 285

\bibitem[{Moffatt(2014)}]{moffatt2014note}
---. 2014, Journal of Fluid Mechanics, 741

\bibitem[{Moffatt(1969)}]{bib:moffatt1969degree}
Moffatt, H.~K. 1969, Journal of Fluid Mechanics, 35, 117

\bibitem[{Moffatt(1978)}]{bib:moffatt1978field}
---. 1978, Cambridge University Press, Cambridge, London, New York, Melbourne

\bibitem[{Monthus(2018)}]{monthus2018various}
Monthus, C. 2018, arXiv preprint arXiv:1810.05082

\bibitem[{M{\"u}ller {et~al.}(2012)M{\"u}ller, Malapaka, \&
  Busse}]{muller2012inverse}
M{\"u}ller, W.-C., Malapaka, S.~K., \& Busse, A. 2012, Physical Review E, 85,
  015302

\bibitem[{Parker(1979)}]{parker1979cosmical}
Parker, E.~N. 1979, Oxford, Clarendon Press; New York, Oxford University Press,
  1979, 858 p.

\bibitem[{Plunian {et~al.}(2013)Plunian, Stepanov, \&
  Frick}]{bib:plunian2013shell}
Plunian, F., Stepanov, R., \& Frick, P. 2013, Physics Reports, 523, 1

\bibitem[{Pouquet {et~al.}(1976)Pouquet, Frisch, \&
  L{\'e}orat}]{bib:pouquet1976strong}
Pouquet, A., Frisch, U., \& L{\'e}orat, J. 1976, Journal of Fluid Mechanics,
  77, 321

\bibitem[{Rathmann \& Ditlevsen(2016)}]{bib:rathmannditlevsen2016}
Rathmann, N.~M., \& Ditlevsen, P.~D. 2016, Phys. Rev. E, 94, 033115,
  \dodoi{10.1103/PhysRevE.94.033115}

\bibitem[{Rathmann \& Ditlevsen(2017)}]{bib:rathmann2017pseudo}
---. 2017, Physical Review Fluids, 2, 054607

\bibitem[{Sahoo {et~al.}(2017{\natexlab{a}})Sahoo, Alexakis, \&
  Biferale}]{bib:sahoo2017discontinuous}
Sahoo, G., Alexakis, A., \& Biferale, L. 2017{\natexlab{a}}, Physical review
  letters, 118, 164501

\bibitem[{Sahoo \& Biferale(2015)}]{bib:sahoo2015disentangling}
Sahoo, G., \& Biferale, L. 2015, The European Physical Journal E, 38, 114

\bibitem[{Sahoo {et~al.}(2015)Sahoo, Bonaccorso, \&
  Biferale}]{bib:sahoo2015roleof}
Sahoo, G., Bonaccorso, F., \& Biferale, L. 2015, Phys. Rev. E, 92, 051002,
  \dodoi{10.1103/PhysRevE.92.051002}

\bibitem[{Sahoo {et~al.}(2017{\natexlab{b}})Sahoo, De~Pietro, \&
  Biferale}]{bib:sahoo2017helicity}
Sahoo, G., De~Pietro, M., \& Biferale, L. 2017{\natexlab{b}}, Physical Review
  Fluids, 2, 024601

\bibitem[{Shukurov(2004)}]{shukurov2004introduction}
Shukurov, A. 2004, arXiv preprint astro-ph/0411739

\bibitem[{Steenbeck {et~al.}(1966)Steenbeck, Krause, \&
  R{\"a}dler}]{steenbeck1966berechnung}
Steenbeck, M., Krause, F., \& R{\"a}dler, K.-H. 1966, Zeitschrift f{\"u}r
  Naturforschung A, 21, 369

\bibitem[{Stepanov {et~al.}(2015)Stepanov, Frick, \& Mizeva}]{bib:Stepanov2015}
Stepanov, R., Frick, P., \& Mizeva, I. 2015, The Astrophysical Journal Letters,
  798, L35

\bibitem[{Sulem {et~al.}(1989)Sulem, She, Scholl, \&
  Frisch}]{bib:sulem1989generation}
Sulem, P., She, Z.~S., Scholl, H., \& Frisch, U. 1989, Journal of Fluid
  Mechanics, 205, 341

\bibitem[{Tobias {et~al.}(2013)Tobias, Cattaneo, Boldyrev, Davidson, Kaneda, \&
  Sreenivasan}]{tobias2013ten}
Tobias, S., Cattaneo, F., Boldyrev, S., {et~al.} 2013, Ten chapters in
  turbulence,  Cambridge University Press Cambridge

\bibitem[{Vainshtein \& Zel'Dovich(1972)}]{vainshtein1972reviews}
Vainshtein, S., \& Zel'Dovich, Y.~B. 1972, Soviet Physics Uspekhi, 15, 159

\bibitem[{Verma {et~al.}(2005)Verma, Ayyer, \& Chandra}]{verma2005energy}
Verma, M.~K., Ayyer, A., \& Chandra, A.~V. 2005, Physics of plasmas, 12, 082307

\bibitem[{Verma \& Kumar(2016)}]{bib:verma2016dynamos}
Verma, M.~K., \& Kumar, R. 2016, Journal of Turbulence, 17, 1112

\bibitem[{Waleffe(1992)}]{bib:waleffe1992nature}
Waleffe, F. 1992, Physics of Fluids A: Fluid Dynamics (1989-1993), 4, 350

\bibitem[{Widrow(2002)}]{widrow2002origin}
Widrow, L.~M. 2002, Reviews of Modern Physics, 74, 775

\end{thebibliography}


\begin{thebibliography}{43}%
\makeatletter
\providecommand \@ifxundefined [1]{%
 \@ifx{#1\undefined}
}%
\providecommand \@ifnum [1]{%
 \ifnum #1\expandafter \@firstoftwo
 \else \expandafter \@secondoftwo
 \fi
}%
\providecommand \@ifx [1]{%
 \ifx #1\expandafter \@firstoftwo
 \else \expandafter \@secondoftwo
 \fi
}%
\providecommand \natexlab [1]{#1}%
\providecommand \enquote  [1]{``#1''}%
\providecommand \bibnamefont  [1]{#1}%
\providecommand \bibfnamefont [1]{#1}%
\providecommand \citenamefont [1]{#1}%
\providecommand \href@noop [0]{\@secondoftwo}%
\providecommand \href [0]{\begingroup \@sanitize@url \@href}%
\providecommand \@href[1]{\@@startlink{#1}\@@href}%
\providecommand \@@href[1]{\endgroup#1\@@endlink}%
\providecommand \@sanitize@url [0]{\catcode `\\12\catcode `\$12\catcode
  `\&12\catcode `\#12\catcode `\^12\catcode `\_12\catcode `\%12\relax}%
\providecommand \@@startlink[1]{}%
\providecommand \@@endlink[0]{}%
\providecommand \url  [0]{\begingroup\@sanitize@url \@url }%
\providecommand \@url [1]{\endgroup\@href {#1}{\urlprefix }}%
\providecommand \urlprefix  [0]{URL }%
\providecommand \Eprint [0]{\href }%
\providecommand \doibase [0]{http://dx.doi.org/}%
\providecommand \selectlanguage [0]{\@gobble}%
\providecommand \bibinfo  [0]{\@secondoftwo}%
\providecommand \bibfield  [0]{\@secondoftwo}%
\providecommand \translation [1]{[#1]}%
\providecommand \BibitemOpen [0]{}%
\providecommand \bibitemStop [0]{}%
\providecommand \bibitemNoStop [0]{.\EOS\space}%
\providecommand \EOS [0]{\spacefactor3000\relax}%
\providecommand \BibitemShut  [1]{\csname bibitem#1\endcsname}%
\let\auto@bib@innerbib\@empty
\bibitem [{\citenamefont {Kraichnan}(1967)}]{bib:kraichnan1967inertial}%
  \BibitemOpen
  \bibfield  {author} {\bibinfo {author} {\bibfnamefont {R.~H.}\ \bibnamefont
  {Kraichnan}},\ }\href@noop {} {\emph {\bibinfo {title} {Inertial ranges in
  two-dimensional turbulence}}},\ \bibinfo {type} {Tech. Rep.}\ (\bibinfo
  {institution} {DTIC Document},\ \bibinfo {year} {1967})\BibitemShut {NoStop}%
\bibitem [{\citenamefont {Moffatt}(1969)}]{bib:moffatt1969degree}%
  \BibitemOpen
  \bibfield  {author} {\bibinfo {author} {\bibfnamefont {H.~K.}\ \bibnamefont
  {Moffatt}},\ }\href@noop {} {\bibfield  {journal} {\bibinfo  {journal}
  {Journal of Fluid Mechanics}\ }\textbf {\bibinfo {volume} {35}},\ \bibinfo
  {pages} {117} (\bibinfo {year} {1969})}\BibitemShut {NoStop}%
\bibitem [{\citenamefont {Kraichnan}(1973)}]{bib:kraichnan1973helical}%
  \BibitemOpen
  \bibfield  {author} {\bibinfo {author} {\bibfnamefont {R.~H.}\ \bibnamefont
  {Kraichnan}},\ }\href@noop {} {\bibfield  {journal} {\bibinfo  {journal}
  {Journal of Fluid Mechanics}\ }\textbf {\bibinfo {volume} {59}},\ \bibinfo
  {pages} {745} (\bibinfo {year} {1973})}\BibitemShut {NoStop}%
\bibitem [{\citenamefont {Brissaud}\ \emph {et~al.}(1973)\citenamefont
  {Brissaud}, \citenamefont {Frisch}, \citenamefont {Leorat}, \citenamefont
  {Lesieur},\ and\ \citenamefont {Mazure}}]{bib:brissaud1973helicity}%
  \BibitemOpen
  \bibfield  {author} {\bibinfo {author} {\bibfnamefont {A.}~\bibnamefont
  {Brissaud}}, \bibinfo {author} {\bibfnamefont {U.}~\bibnamefont {Frisch}},
  \bibinfo {author} {\bibfnamefont {J.}~\bibnamefont {Leorat}}, \bibinfo
  {author} {\bibfnamefont {M.}~\bibnamefont {Lesieur}}, \ and\ \bibinfo
  {author} {\bibfnamefont {A.}~\bibnamefont {Mazure}},\ }\href@noop {}
  {\bibfield  {journal} {\bibinfo  {journal} {Physics of Fluids (1958-1988)}\
  }\textbf {\bibinfo {volume} {16}},\ \bibinfo {pages} {1366} (\bibinfo {year}
  {1973})}\BibitemShut {NoStop}%
\bibitem [{\citenamefont {Ditlevsen}\ and\ \citenamefont
  {Giuliani}(2001{\natexlab{a}})}]{bib:ditlevsen2001dissipation}%
  \BibitemOpen
  \bibfield  {author} {\bibinfo {author} {\bibfnamefont {P.~D.}\ \bibnamefont
  {Ditlevsen}}\ and\ \bibinfo {author} {\bibfnamefont {P.}~\bibnamefont
  {Giuliani}},\ }\href@noop {} {\bibfield  {journal} {\bibinfo  {journal}
  {Physics of Fluids (1994-present)}\ }\textbf {\bibinfo {volume} {13}},\
  \bibinfo {pages} {3508} (\bibinfo {year} {2001}{\natexlab{a}})}\BibitemShut
  {NoStop}%
\bibitem [{\citenamefont {Sulem}\ \emph {et~al.}(1989)\citenamefont {Sulem},
  \citenamefont {She}, \citenamefont {Scholl},\ and\ \citenamefont
  {Frisch}}]{bib:sulem1989generation}%
  \BibitemOpen
  \bibfield  {author} {\bibinfo {author} {\bibfnamefont {P.}~\bibnamefont
  {Sulem}}, \bibinfo {author} {\bibfnamefont {Z.~S.}\ \bibnamefont {She}},
  \bibinfo {author} {\bibfnamefont {H.}~\bibnamefont {Scholl}}, \ and\ \bibinfo
  {author} {\bibfnamefont {U.}~\bibnamefont {Frisch}},\ }\href@noop {}
  {\bibfield  {journal} {\bibinfo  {journal} {Journal of Fluid Mechanics}\
  }\textbf {\bibinfo {volume} {205}},\ \bibinfo {pages} {341} (\bibinfo {year}
  {1989})}\BibitemShut {NoStop}%
\bibitem [{\citenamefont {Mininni}\ \emph {et~al.}(2009)\citenamefont
  {Mininni}, \citenamefont {Alexakis},\ and\ \citenamefont
  {Pouquet}}]{bib:mininni2009scale}%
  \BibitemOpen
  \bibfield  {author} {\bibinfo {author} {\bibfnamefont {P.~D.}\ \bibnamefont
  {Mininni}}, \bibinfo {author} {\bibfnamefont {A.}~\bibnamefont {Alexakis}}, \
  and\ \bibinfo {author} {\bibfnamefont {A.}~\bibnamefont {Pouquet}},\
  }\href@noop {} {\bibfield  {journal} {\bibinfo  {journal} {Physics of Fluids
  (1994-present)}\ }\textbf {\bibinfo {volume} {21}},\ \bibinfo {pages}
  {015108} (\bibinfo {year} {2009})}\BibitemShut {NoStop}%
\bibitem [{\citenamefont {Constantin}\ and\ \citenamefont
  {Majda}(1988)}]{bib:constantin1988beltrami}%
  \BibitemOpen
  \bibfield  {author} {\bibinfo {author} {\bibfnamefont {P.}~\bibnamefont
  {Constantin}}\ and\ \bibinfo {author} {\bibfnamefont {A.}~\bibnamefont
  {Majda}},\ }\href@noop {} {\bibfield  {journal} {\bibinfo  {journal}
  {Communications in mathematical physics}\ }\textbf {\bibinfo {volume}
  {115}},\ \bibinfo {pages} {435} (\bibinfo {year} {1988})}\BibitemShut
  {NoStop}%
\bibitem [{\citenamefont {Waleffe}(1992)}]{bib:waleffe1992nature}%
  \BibitemOpen
  \bibfield  {author} {\bibinfo {author} {\bibfnamefont {F.}~\bibnamefont
  {Waleffe}},\ }\href@noop {} {\bibfield  {journal} {\bibinfo  {journal}
  {Physics of Fluids A: Fluid Dynamics (1989-1993)}\ }\textbf {\bibinfo
  {volume} {4}},\ \bibinfo {pages} {350} (\bibinfo {year} {1992})}\BibitemShut
  {NoStop}%
\bibitem [{\citenamefont {Biferale}\ \emph {et~al.}(2012)\citenamefont
  {Biferale}, \citenamefont {Musacchio},\ and\ \citenamefont
  {Toschi}}]{bib:biferale2012inverse}%
  \BibitemOpen
  \bibfield  {author} {\bibinfo {author} {\bibfnamefont {L.}~\bibnamefont
  {Biferale}}, \bibinfo {author} {\bibfnamefont {S.}~\bibnamefont {Musacchio}},
  \ and\ \bibinfo {author} {\bibfnamefont {F.}~\bibnamefont {Toschi}},\
  }\href@noop {} {\bibfield  {journal} {\bibinfo  {journal} {Physical review
  letters}\ }\textbf {\bibinfo {volume} {108}},\ \bibinfo {pages} {164501}
  (\bibinfo {year} {2012})}\BibitemShut {NoStop}%
\bibitem [{\citenamefont {Herbert}(2014)}]{bib:herbert2014restricted}%
  \BibitemOpen
  \bibfield  {author} {\bibinfo {author} {\bibfnamefont {C.}~\bibnamefont
  {Herbert}},\ }\href@noop {} {\bibfield  {journal} {\bibinfo  {journal}
  {Physical Review E}\ }\textbf {\bibinfo {volume} {89}},\ \bibinfo {pages}
  {013010} (\bibinfo {year} {2014})}\BibitemShut {NoStop}%
\bibitem [{\citenamefont {Rathmann}\ and\ \citenamefont
  {Ditlevsen}(2017)}]{bib:rathmann2017pseudo}%
  \BibitemOpen
  \bibfield  {author} {\bibinfo {author} {\bibfnamefont {N.~M.}\ \bibnamefont
  {Rathmann}}\ and\ \bibinfo {author} {\bibfnamefont {P.~D.}\ \bibnamefont
  {Ditlevsen}},\ }\href@noop {} {\bibfield  {journal} {\bibinfo  {journal}
  {Physical Review Fluids}\ }\textbf {\bibinfo {volume} {2}},\ \bibinfo {pages}
  {054607} (\bibinfo {year} {2017})}\BibitemShut {NoStop}%
\bibitem [{\citenamefont {De~Pietro}\ \emph {et~al.}(2015)\citenamefont
  {De~Pietro}, \citenamefont {Biferale},\ and\ \citenamefont
  {Mailybaev}}]{bib:DePietro2015}%
  \BibitemOpen
  \bibfield  {author} {\bibinfo {author} {\bibfnamefont {M.}~\bibnamefont
  {De~Pietro}}, \bibinfo {author} {\bibfnamefont {L.}~\bibnamefont {Biferale}},
  \ and\ \bibinfo {author} {\bibfnamefont {A.~A.}\ \bibnamefont {Mailybaev}},\
  }\href {\doibase 10.1103/PhysRevE.92.043021} {\bibfield  {journal} {\bibinfo
  {journal} {Phys. Rev. E}\ }\textbf {\bibinfo {volume} {92}},\ \bibinfo
  {pages} {043021} (\bibinfo {year} {2015})}\BibitemShut {NoStop}%
\bibitem [{\citenamefont {Musacchio}\ and\ \citenamefont
  {Boffetta}(2017)}]{musacchio2017split}%
  \BibitemOpen
  \bibfield  {author} {\bibinfo {author} {\bibfnamefont {S.}~\bibnamefont
  {Musacchio}}\ and\ \bibinfo {author} {\bibfnamefont {G.}~\bibnamefont
  {Boffetta}},\ }\href@noop {} {\bibfield  {journal} {\bibinfo  {journal}
  {Physics of Fluids}\ }\textbf {\bibinfo {volume} {29}},\ \bibinfo {pages}
  {111106} (\bibinfo {year} {2017})}\BibitemShut {NoStop}%
\bibitem [{\citenamefont {Pouquet}\ \emph {et~al.}(2017)\citenamefont
  {Pouquet}, \citenamefont {Marino}, \citenamefont {Mininni},\ and\
  \citenamefont {Rosenberg}}]{bib:pouquet2017dual}%
  \BibitemOpen
  \bibfield  {author} {\bibinfo {author} {\bibfnamefont {A.}~\bibnamefont
  {Pouquet}}, \bibinfo {author} {\bibfnamefont {R.}~\bibnamefont {Marino}},
  \bibinfo {author} {\bibfnamefont {P.~D.}\ \bibnamefont {Mininni}}, \ and\
  \bibinfo {author} {\bibfnamefont {D.}~\bibnamefont {Rosenberg}},\ }\href@noop
  {} {\bibfield  {journal} {\bibinfo  {journal} {Physics of Fluids}\ }\textbf
  {\bibinfo {volume} {29}},\ \bibinfo {pages} {111108} (\bibinfo {year}
  {2017})}\BibitemShut {NoStop}%
\bibitem [{\citenamefont {Buzzicotti}\ \emph {et~al.}(2017)\citenamefont
  {Buzzicotti}, \citenamefont {Aluie}, \citenamefont {Biferale},\ and\
  \citenamefont {Linkmann}}]{bib:buzzicotti2017energy}%
  \BibitemOpen
  \bibfield  {author} {\bibinfo {author} {\bibfnamefont {M.}~\bibnamefont
  {Buzzicotti}}, \bibinfo {author} {\bibfnamefont {H.}~\bibnamefont {Aluie}},
  \bibinfo {author} {\bibfnamefont {L.}~\bibnamefont {Biferale}}, \ and\
  \bibinfo {author} {\bibfnamefont {M.}~\bibnamefont {Linkmann}},\ }\href@noop
  {} {\bibfield  {journal} {\bibinfo  {journal} {arXiv preprint
  arXiv:1711.07054}\ } (\bibinfo {year} {2017})}\BibitemShut {NoStop}%
\bibitem [{\citenamefont {Benavides}\ and\ \citenamefont
  {Alexakis}(2017)}]{bib:benavides2017critical}%
  \BibitemOpen
  \bibfield  {author} {\bibinfo {author} {\bibfnamefont {S.~J.}\ \bibnamefont
  {Benavides}}\ and\ \bibinfo {author} {\bibfnamefont {A.}~\bibnamefont
  {Alexakis}},\ }\href@noop {} {\bibfield  {journal} {\bibinfo  {journal}
  {Journal of Fluid Mechanics}\ }\textbf {\bibinfo {volume} {822}},\ \bibinfo
  {pages} {364} (\bibinfo {year} {2017})}\BibitemShut {NoStop}%
\bibitem [{\citenamefont {Qu}\ \emph {et~al.}(2018)\citenamefont {Qu},
  \citenamefont {Naso},\ and\ \citenamefont {Bos}}]{bib:qu2018cascades}%
  \BibitemOpen
  \bibfield  {author} {\bibinfo {author} {\bibfnamefont {B.}~\bibnamefont
  {Qu}}, \bibinfo {author} {\bibfnamefont {A.}~\bibnamefont {Naso}}, \ and\
  \bibinfo {author} {\bibfnamefont {W.~J.}\ \bibnamefont {Bos}},\ }\href@noop
  {} {\bibfield  {journal} {\bibinfo  {journal} {Physical Review Fluids}\
  }\textbf {\bibinfo {volume} {3}},\ \bibinfo {pages} {014607} (\bibinfo {year}
  {2018})}\BibitemShut {NoStop}%
\bibitem [{\citenamefont {Biferale}\ \emph {et~al.}(2013)\citenamefont
  {Biferale}, \citenamefont {Musacchio},\ and\ \citenamefont
  {Toschi}}]{bib:biferale2013split}%
  \BibitemOpen
  \bibfield  {author} {\bibinfo {author} {\bibfnamefont {L.}~\bibnamefont
  {Biferale}}, \bibinfo {author} {\bibfnamefont {S.}~\bibnamefont {Musacchio}},
  \ and\ \bibinfo {author} {\bibfnamefont {F.}~\bibnamefont {Toschi}},\
  }\href@noop {} {\bibfield  {journal} {\bibinfo  {journal} {Journal of Fluid
  Mechanics}\ }\textbf {\bibinfo {volume} {730}},\ \bibinfo {pages} {309}
  (\bibinfo {year} {2013})}\BibitemShut {NoStop}%
\bibitem [{\citenamefont {Sahoo}\ \emph {et~al.}(2015)\citenamefont {Sahoo},
  \citenamefont {Bonaccorso},\ and\ \citenamefont {Biferale}}]{bib:sahoo2015}%
  \BibitemOpen
  \bibfield  {author} {\bibinfo {author} {\bibfnamefont {G.}~\bibnamefont
  {Sahoo}}, \bibinfo {author} {\bibfnamefont {F.}~\bibnamefont {Bonaccorso}}, \
  and\ \bibinfo {author} {\bibfnamefont {L.}~\bibnamefont {Biferale}},\ }\href
  {\doibase 10.1103/PhysRevE.92.051002} {\bibfield  {journal} {\bibinfo
  {journal} {Phys. Rev. E}\ }\textbf {\bibinfo {volume} {92}},\ \bibinfo
  {pages} {051002} (\bibinfo {year} {2015})}\BibitemShut {NoStop}%
\bibitem [{\citenamefont {Sahoo}\ and\ \citenamefont
  {Biferale}(2015)}]{bib:sahoo2015disentangling}%
  \BibitemOpen
  \bibfield  {author} {\bibinfo {author} {\bibfnamefont {G.}~\bibnamefont
  {Sahoo}}\ and\ \bibinfo {author} {\bibfnamefont {L.}~\bibnamefont
  {Biferale}},\ }\href@noop {} {\bibfield  {journal} {\bibinfo  {journal} {The
  European Physical Journal E}\ }\textbf {\bibinfo {volume} {38}},\ \bibinfo
  {pages} {114} (\bibinfo {year} {2015})}\BibitemShut {NoStop}%
\bibitem [{\citenamefont {Rathmann}\ and\ \citenamefont
  {Ditlevsen}(2016)}]{bib:rathmannditlevsen2016}%
  \BibitemOpen
  \bibfield  {author} {\bibinfo {author} {\bibfnamefont {N.~M.}\ \bibnamefont
  {Rathmann}}\ and\ \bibinfo {author} {\bibfnamefont {P.~D.}\ \bibnamefont
  {Ditlevsen}},\ }\href {\doibase 10.1103/PhysRevE.94.033115} {\bibfield
  {journal} {\bibinfo  {journal} {Phys. Rev. E}\ }\textbf {\bibinfo {volume}
  {94}},\ \bibinfo {pages} {033115} (\bibinfo {year} {2016})}\BibitemShut
  {NoStop}%
\bibitem [{\citenamefont {Alexakis}(2017)}]{bib:alexakis2016helically}%
  \BibitemOpen
  \bibfield  {author} {\bibinfo {author} {\bibfnamefont {A.}~\bibnamefont
  {Alexakis}},\ }\href@noop {} {\bibfield  {journal} {\bibinfo  {journal}
  {Journal of Fluid Mechanics}\ }\textbf {\bibinfo {volume} {812}},\ \bibinfo
  {pages} {752} (\bibinfo {year} {2017})}\BibitemShut {NoStop}%
\bibitem [{\citenamefont {Sahoo}\ \emph
  {et~al.}(2017{\natexlab{a}})\citenamefont {Sahoo}, \citenamefont {Alexakis},\
  and\ \citenamefont {Biferale}}]{bib:sahoo2017discontinuous}%
  \BibitemOpen
  \bibfield  {author} {\bibinfo {author} {\bibfnamefont {G.}~\bibnamefont
  {Sahoo}}, \bibinfo {author} {\bibfnamefont {A.}~\bibnamefont {Alexakis}}, \
  and\ \bibinfo {author} {\bibfnamefont {L.}~\bibnamefont {Biferale}},\
  }\href@noop {} {\bibfield  {journal} {\bibinfo  {journal} {Physical review
  letters}\ }\textbf {\bibinfo {volume} {118}},\ \bibinfo {pages} {164501}
  (\bibinfo {year} {2017}{\natexlab{a}})}\BibitemShut {NoStop}%
\bibitem [{\citenamefont {Biferale}\ \emph {et~al.}(2017)\citenamefont
  {Biferale}, \citenamefont {Buzzicotti},\ and\ \citenamefont
  {Linkmann}}]{bib:biferale2017two}%
  \BibitemOpen
  \bibfield  {author} {\bibinfo {author} {\bibfnamefont {L.}~\bibnamefont
  {Biferale}}, \bibinfo {author} {\bibfnamefont {M.}~\bibnamefont
  {Buzzicotti}}, \ and\ \bibinfo {author} {\bibfnamefont {M.}~\bibnamefont
  {Linkmann}},\ }\href@noop {} {\bibfield  {journal} {\bibinfo  {journal}
  {Physics of Fluids}\ }\textbf {\bibinfo {volume} {29}},\ \bibinfo {pages}
  {111101} (\bibinfo {year} {2017})}\BibitemShut {NoStop}%
\bibitem [{\citenamefont {Sahoo}\ and\ \citenamefont
  {Biferale}(2018)}]{bib:sahoo2018energy}%
  \BibitemOpen
  \bibfield  {author} {\bibinfo {author} {\bibfnamefont {G.}~\bibnamefont
  {Sahoo}}\ and\ \bibinfo {author} {\bibfnamefont {L.}~\bibnamefont
  {Biferale}},\ }\href@noop {} {\bibfield  {journal} {\bibinfo  {journal}
  {Fluid Dynamics Research}\ }\textbf {\bibinfo {volume} {50}},\ \bibinfo
  {pages} {011420} (\bibinfo {year} {2018})}\BibitemShut {NoStop}%
\bibitem [{\citenamefont {Lessinnes}\ \emph {et~al.}(2009)\citenamefont
  {Lessinnes}, \citenamefont {Plunian},\ and\ \citenamefont
  {Carati}}]{bib:lessinnes2009helical}%
  \BibitemOpen
  \bibfield  {author} {\bibinfo {author} {\bibfnamefont {T.}~\bibnamefont
  {Lessinnes}}, \bibinfo {author} {\bibfnamefont {F.}~\bibnamefont {Plunian}},
  \ and\ \bibinfo {author} {\bibfnamefont {D.}~\bibnamefont {Carati}},\
  }\href@noop {} {\bibfield  {journal} {\bibinfo  {journal} {Theoretical and
  Computational Fluid Dynamics}\ }\textbf {\bibinfo {volume} {23}},\ \bibinfo
  {pages} {439} (\bibinfo {year} {2009})}\BibitemShut {NoStop}%
\bibitem [{\citenamefont {Linkmann}\ \emph {et~al.}(2017)\citenamefont
  {Linkmann}, \citenamefont {Sahoo}, \citenamefont {McKay}, \citenamefont
  {Berera},\ and\ \citenamefont {Biferale}}]{bib:linkmann2017effects}%
  \BibitemOpen
  \bibfield  {author} {\bibinfo {author} {\bibfnamefont {M.}~\bibnamefont
  {Linkmann}}, \bibinfo {author} {\bibfnamefont {G.}~\bibnamefont {Sahoo}},
  \bibinfo {author} {\bibfnamefont {M.}~\bibnamefont {McKay}}, \bibinfo
  {author} {\bibfnamefont {A.}~\bibnamefont {Berera}}, \ and\ \bibinfo {author}
  {\bibfnamefont {L.}~\bibnamefont {Biferale}},\ }\href@noop {} {\bibfield
  {journal} {\bibinfo  {journal} {The Astrophysical Journal}\ }\textbf
  {\bibinfo {volume} {836}},\ \bibinfo {pages} {26} (\bibinfo {year}
  {2017})}\BibitemShut {NoStop}%
\bibitem [{\citenamefont {Iroshnikov}(1964)}]{bib:iroshnikov1964turbulence}%
  \BibitemOpen
  \bibfield  {author} {\bibinfo {author} {\bibfnamefont {P.}~\bibnamefont
  {Iroshnikov}},\ }\href@noop {} {\bibfield  {journal} {\bibinfo  {journal}
  {Soviet Astronomy}\ }\textbf {\bibinfo {volume} {7}},\ \bibinfo {pages} {566}
  (\bibinfo {year} {1964})}\BibitemShut {NoStop}%
\bibitem [{\citenamefont {Matthaeus}\ and\ \citenamefont
  {Zhou}(1989)}]{bib:matthaeus1989extended}%
  \BibitemOpen
  \bibfield  {author} {\bibinfo {author} {\bibfnamefont {W.~H.}\ \bibnamefont
  {Matthaeus}}\ and\ \bibinfo {author} {\bibfnamefont {Y.}~\bibnamefont
  {Zhou}},\ }\href@noop {} {\bibfield  {journal} {\bibinfo  {journal} {Physics
  of Fluids B: Plasma Physics}\ }\textbf {\bibinfo {volume} {1}},\ \bibinfo
  {pages} {1929} (\bibinfo {year} {1989})}\BibitemShut {NoStop}%
\bibitem [{\citenamefont {Linkmann}\ and\ \citenamefont
  {Dallas}(2017)}]{bib:linkmann2017triad}%
  \BibitemOpen
  \bibfield  {author} {\bibinfo {author} {\bibfnamefont {M.}~\bibnamefont
  {Linkmann}}\ and\ \bibinfo {author} {\bibfnamefont {V.}~\bibnamefont
  {Dallas}},\ }\href@noop {} {\bibfield  {journal} {\bibinfo  {journal}
  {Physical Review Fluids}\ }\textbf {\bibinfo {volume} {2}},\ \bibinfo {pages}
  {054605} (\bibinfo {year} {2017})}\BibitemShut {NoStop}%
\bibitem [{\citenamefont {Linkmann}\ \emph {et~al.}(2016)\citenamefont
  {Linkmann}, \citenamefont {Berera}, \citenamefont {McKay},\ and\
  \citenamefont {J{\"a}ger}}]{bib:linkmann2016helical}%
  \BibitemOpen
  \bibfield  {author} {\bibinfo {author} {\bibfnamefont {M.}~\bibnamefont
  {Linkmann}}, \bibinfo {author} {\bibfnamefont {A.}~\bibnamefont {Berera}},
  \bibinfo {author} {\bibfnamefont {M.}~\bibnamefont {McKay}}, \ and\ \bibinfo
  {author} {\bibfnamefont {J.}~\bibnamefont {J{\"a}ger}},\ }\href@noop {}
  {\bibfield  {journal} {\bibinfo  {journal} {Journal of Fluid Mechanics}\
  }\textbf {\bibinfo {volume} {791}},\ \bibinfo {pages} {61} (\bibinfo {year}
  {2016})}\BibitemShut {NoStop}%
\bibitem [{\citenamefont {Benzi}\ \emph {et~al.}(1996)\citenamefont {Benzi},
  \citenamefont {Biferale}, \citenamefont {Kerr},\ and\ \citenamefont
  {Trovatore}}]{bib:benzi1996helical}%
  \BibitemOpen
  \bibfield  {author} {\bibinfo {author} {\bibfnamefont {R.}~\bibnamefont
  {Benzi}}, \bibinfo {author} {\bibfnamefont {L.}~\bibnamefont {Biferale}},
  \bibinfo {author} {\bibfnamefont {R.}~\bibnamefont {Kerr}}, \ and\ \bibinfo
  {author} {\bibfnamefont {E.}~\bibnamefont {Trovatore}},\ }\href@noop {}
  {\bibfield  {journal} {\bibinfo  {journal} {Physical Review E}\ }\textbf
  {\bibinfo {volume} {53}},\ \bibinfo {pages} {3541} (\bibinfo {year}
  {1996})}\BibitemShut {NoStop}%
\bibitem [{\citenamefont {Ditlevsen}\ and\ \citenamefont
  {Giuliani}(2001{\natexlab{b}})}]{bib:ditlevsen2001cascades}%
  \BibitemOpen
  \bibfield  {author} {\bibinfo {author} {\bibfnamefont {P.~D.}\ \bibnamefont
  {Ditlevsen}}\ and\ \bibinfo {author} {\bibfnamefont {P.}~\bibnamefont
  {Giuliani}},\ }\href@noop {} {\bibfield  {journal} {\bibinfo  {journal}
  {Physical Review E}\ }\textbf {\bibinfo {volume} {63}},\ \bibinfo {pages}
  {036304} (\bibinfo {year} {2001}{\natexlab{b}})}\BibitemShut {NoStop}%
\bibitem [{\citenamefont {Lessinnes}\ \emph {et~al.}(2011)\citenamefont
  {Lessinnes}, \citenamefont {Plunian}, \citenamefont {Stepanov},\ and\
  \citenamefont {Carati}}]{bib:lessinnes2011dissipation}%
  \BibitemOpen
  \bibfield  {author} {\bibinfo {author} {\bibfnamefont {T.}~\bibnamefont
  {Lessinnes}}, \bibinfo {author} {\bibfnamefont {F.}~\bibnamefont {Plunian}},
  \bibinfo {author} {\bibfnamefont {R.}~\bibnamefont {Stepanov}}, \ and\
  \bibinfo {author} {\bibfnamefont {D.}~\bibnamefont {Carati}},\ }\href@noop {}
  {\bibfield  {journal} {\bibinfo  {journal} {Physics of Fluids
  (1994-present)}\ }\textbf {\bibinfo {volume} {23}},\ \bibinfo {pages}
  {035108} (\bibinfo {year} {2011})}\BibitemShut {NoStop}%
\bibitem [{\citenamefont {Stepanov}\ \emph {et~al.}(2015)\citenamefont
  {Stepanov}, \citenamefont {Frick},\ and\ \citenamefont
  {Mizeva}}]{bib:Stepanov2015}%
  \BibitemOpen
  \bibfield  {author} {\bibinfo {author} {\bibfnamefont {R.}~\bibnamefont
  {Stepanov}}, \bibinfo {author} {\bibfnamefont {P.}~\bibnamefont {Frick}}, \
  and\ \bibinfo {author} {\bibfnamefont {I.}~\bibnamefont {Mizeva}},\ }\href
  {http://stacks.iop.org/2041-8205/798/i=2/a=L35} {\bibfield  {journal}
  {\bibinfo  {journal} {The Astrophysical Journal Letters}\ }\textbf {\bibinfo
  {volume} {798}},\ \bibinfo {pages} {L35} (\bibinfo {year}
  {2015})}\BibitemShut {NoStop}%
\bibitem [{\citenamefont {Sahoo}\ \emph
  {et~al.}(2017{\natexlab{b}})\citenamefont {Sahoo}, \citenamefont
  {De~Pietro},\ and\ \citenamefont {Biferale}}]{bib:sahoo2017helicity}%
  \BibitemOpen
  \bibfield  {author} {\bibinfo {author} {\bibfnamefont {G.}~\bibnamefont
  {Sahoo}}, \bibinfo {author} {\bibfnamefont {M.}~\bibnamefont {De~Pietro}}, \
  and\ \bibinfo {author} {\bibfnamefont {L.}~\bibnamefont {Biferale}},\
  }\href@noop {} {\bibfield  {journal} {\bibinfo  {journal} {Physical Review
  Fluids}\ }\textbf {\bibinfo {volume} {2}},\ \bibinfo {pages} {024601}
  (\bibinfo {year} {2017}{\natexlab{b}})}\BibitemShut {NoStop}%
\bibitem [{\citenamefont {De~Pietro}\ \emph {et~al.}(2017)\citenamefont
  {De~Pietro}, \citenamefont {Mailybaev},\ and\ \citenamefont
  {Biferale}}]{bib:DePietro2016chaotic}%
  \BibitemOpen
  \bibfield  {author} {\bibinfo {author} {\bibfnamefont {M.}~\bibnamefont
  {De~Pietro}}, \bibinfo {author} {\bibfnamefont {A.~A.}\ \bibnamefont
  {Mailybaev}}, \ and\ \bibinfo {author} {\bibfnamefont {L.}~\bibnamefont
  {Biferale}},\ }\href@noop {} {\bibfield  {journal} {\bibinfo  {journal}
  {Physical Review Fluids}\ }\textbf {\bibinfo {volume} {2}},\ \bibinfo {pages}
  {034606} (\bibinfo {year} {2017})}\BibitemShut {NoStop}%
\bibitem [{\citenamefont {Plunian}\ \emph {et~al.}(2013)\citenamefont
  {Plunian}, \citenamefont {Stepanov},\ and\ \citenamefont
  {Frick}}]{bib:plunian2013shell}%
  \BibitemOpen
  \bibfield  {author} {\bibinfo {author} {\bibfnamefont {F.}~\bibnamefont
  {Plunian}}, \bibinfo {author} {\bibfnamefont {R.}~\bibnamefont {Stepanov}}, \
  and\ \bibinfo {author} {\bibfnamefont {P.}~\bibnamefont {Frick}},\
  }\href@noop {} {\bibfield  {journal} {\bibinfo  {journal} {Physics Reports}\
  }\textbf {\bibinfo {volume} {523}},\ \bibinfo {pages} {1} (\bibinfo {year}
  {2013})}\BibitemShut {NoStop}%
\bibitem [{\citenamefont {Verma}\ and\ \citenamefont
  {Kumar}(2016)}]{bib:verma2016dynamos}%
  \BibitemOpen
  \bibfield  {author} {\bibinfo {author} {\bibfnamefont {M.~K.}\ \bibnamefont
  {Verma}}\ and\ \bibinfo {author} {\bibfnamefont {R.}~\bibnamefont {Kumar}},\
  }\href@noop {} {\bibfield  {journal} {\bibinfo  {journal} {Journal of
  Turbulence}\ }\textbf {\bibinfo {volume} {17}},\ \bibinfo {pages} {1112}
  (\bibinfo {year} {2016})}\BibitemShut {NoStop}%
\bibitem [{\citenamefont {Moffatt}(1978)}]{bib:moffatt1978field}%
  \BibitemOpen
  \bibfield  {author} {\bibinfo {author} {\bibfnamefont {H.~K.}\ \bibnamefont
  {Moffatt}},\ }\href@noop {} {\bibfield  {journal} {\bibinfo  {journal}
  {Cambridge University Press, Cambridge, London, New York, Melbourne}\ }
  (\bibinfo {year} {1978})}\BibitemShut {NoStop}%
\bibitem [{\citenamefont {Pouquet}\ \emph {et~al.}(1976)\citenamefont
  {Pouquet}, \citenamefont {Frisch},\ and\ \citenamefont
  {L{\'e}orat}}]{bib:pouquet1976strong}%
  \BibitemOpen
  \bibfield  {author} {\bibinfo {author} {\bibfnamefont {A.}~\bibnamefont
  {Pouquet}}, \bibinfo {author} {\bibfnamefont {U.}~\bibnamefont {Frisch}}, \
  and\ \bibinfo {author} {\bibfnamefont {J.}~\bibnamefont {L{\'e}orat}},\
  }\href@noop {} {\bibfield  {journal} {\bibinfo  {journal} {Journal of Fluid
  Mechanics}\ }\textbf {\bibinfo {volume} {77}},\ \bibinfo {pages} {321}
  (\bibinfo {year} {1976})}\BibitemShut {NoStop}%
\bibitem [{\citenamefont {Biskamp}(1993)}]{bib:biskamp1993nonlinear}%
  \BibitemOpen
  \bibfield  {author} {\bibinfo {author} {\bibfnamefont {D.}~\bibnamefont
  {Biskamp}},\ }\href@noop {} {\enquote {\bibinfo {title} {Nonlinear
  magnetohydrodynamics, volume 1 of cambridge monographs on plasma physics},}\
  } (\bibinfo {year} {1993})\BibitemShut {NoStop}%
\end{thebibliography}%

\end{document}